\documentclass[showpacs,preprintnumbers,amsmath,amssymb,nofootinbib]{revtex4}
\usepackage{graphicx}
\usepackage{dcolumn}
\usepackage{bm}
\begin{document}


\title{Modified Hartree-Fock-Bogoliubov theory at finite 
temperature\footnote{to be published in Physical Review C}}

\author{Nguyen Dinh Dang}
\altaffiliation[Also at ]
{Institute for Nuclear Science and Technique, VAEC, Hanoi, Vietnam}
\affiliation{%
RI-beam factory project office, RIKEN, 2-1 Hirosawa, Wako, 351-0198 
 Saitama, Japan}%
\author{Akito Arima}
 
\affiliation{House of Councilors, 2-1-1 Nagata-cho, 
Chiyoda-ku, Tokyo 100-8962, Japan}%

\date{May 3, 2003}
\begin{abstract}
The modified HFB (MHFB) theory at finite temperature is
derived, which conserves the unitarity relation of the 
particle-density matrix.
This is achieved by constructing a modified quasiparticle-density 
matrix, where the fluctuation of the quasiparticle number is 
microscopically built in. This matrix can be directly obtained from 
the usual 
quasiparticle-density matrix by applying the secondary Bogoliubov 
transformation, 
which includes the quasiparticle occupation number. 
It is shown that, in the limit of constant pairing parameter, 
the MHFB theory yields
the previously obtained modified BCS (MBCS) equations. 
It is also proved that the modified 
quasiparticle RPA, which is based on 
the MBCS quasiparticle excitations, 
conserves the Ikeda sum rule. 
The numerical calculations of the pairing gap, heat capacity, level 
density, and level density 
parameter within the MBCS theory are carried out for $^{120}$Sn.
The results show that 
the superfluid - normal phase transition 
is completely washed out. The applicability of the 
MBCS up to a temperature as high as $T\sim$ 5 MeV is analyzed in 
detail.
\end{abstract}
\pacs{PACS numbers: 21.10.Ma, 21.60.Jz, 24.60.-k, 
27.60.+j }
\maketitle
\section{Introduction}
The finite-temperature Hartree-Fock-Bogoliubov (FT-HFB) theory has 
been 
successfully applied to highly excited nuclei~\cite{Goodman1,Tanabe}. 
It offers a fully 
self-consistent treatment of the 
interplay between single-particle, pairing as well as 
rotational degrees of freedom for nuclei in thermal 
equilibrium.

A major drawback of this theory is the omission of fluctuation 
effects, which can be classified as quantal and statistical 
fluctuations. Quantal fluctuations arise from the mean-field 
approximation
to the exact density operator ${\cal D}$. As a result,
the Hartree-Fock-Bogoliubov (HFB) 
density operator ${\cal D}_{\rm HFB}$ violates the symmetries
of the single-particle Hamiltonian $H$ such as the conservation of 
particle number and spin. However, quantal fluctuations decrease
as the temperature increases. Various methods such as the 
Lipkin-Nogami~\cite{LN} 
method, particle-number projection~\cite{Sheikh}, angular-momentum 
projection
~\cite{RingSchuck}, particle-number conserving 
pairing correlations~\cite{Quentin}, etc. 
have been proposed to eliminate quantal fluctuations. 

On the contrary, statistical fluctuations, which
appear at finite temperature ($T\neq$ 0),
increase with increasing $T$~\cite{Egido,DangZ,Goodman2}. 
Even the knowledge of the exact density operator ${\cal D}$ 
would not eliminate
statistical fluctuations from the FT-HFB theory. 
The omission of statistical fluctuation 
effect leads to the violation of another symmetry, namely the 
unitarity relation of the particle-density 
matrix~\cite{Goodman1,Goodman2}. 
An immediate consequence of this symmetry violation is 
the collapse of the pairing gap $\Delta(T)$ at a critical 
temperature $T_{\rm c}\approx\frac{1}{2}\Delta(T=0)$ in all 
calculations for realistic nuclei 
within the FT-HFB theory and its limit, the finite-temperature BCS 
(FT-BCS)
theory{\cite{Goodman1,Tanabe,Goodman2,Sandu}. 
Such collapse of the pairing gap has been 
usually speculated as the signature of the superfuid-normal phase 
transition in finite nuclei. However, by using the Landau 
macroscopic theory of phase transition~\cite{Landau}, 
Moretto has shown a long time
ago that statistical fluctuations 
wash out such phase transition in finite systems such as nuclei, 
where these fluctuations are indeed quite large~\cite{Moretto}. 
This conclusion has been confirmed recently
by the calculations within the 
modified BCS (MBCS) theory~\cite{MRPA,MBCS}. 
The latter employs the modified quasiparticles obtained by a 
secondary 
Bogoliubov transformation of usual quasiparticles explicitly 
involving the 
quasiparticle occupation numbers. Other approaches such as the 
static-path approximation~\cite{DangRing,Rossignoli}, shell-model 
Monte-Carlo approach~\cite{Liu}, modern nuclear 
shell model 
calculations~\cite{Ze}, 
as well as the exact solution of the pairing problem 
\cite{Volya} also show that pairing 
correlations do not abruptly disappear at $T\neq$ 0.

Another example of symmetry violation caused by the HFB 
and/or BCS theories is the violation of the Ikeda sum rule within the 
renormalized quasiparticle random-phase approximation (renormalized 
QRPA).
The Ikeda sum rule states that the difference ${\cal S}^{-}-{\cal 
S}^{+}
=(2J+1)(N-Z)$ 
between the total strength
${\cal S}^{-}$ of $\beta^{-}$ transitions and of $\beta^{+}$ ones, 
${\cal S}^{+}$, 
is independent of models, where 
$N$ and $Z$ 
are the neutron and proton numbers, respectively, and 
$J$ is the angular momentum of the transitions~\cite{Ikeda}. 
The renormalized RPA (or the renormalized QRPA, which 
includes pairing correlations), is an approach to take into account 
the Pauli principle between the particle (quasiparticle) pairs, 
which the RPA (QRPA) ignores
~\cite{Hara,RRPA,Krmpotic}. This renormalises the RPA 
forward-going $X$ and backward-going $Y$ amplitudes as well as the 
two-body interaction matrix elements by a factor, 
which involves the particle (quasiparticle) occupation numbers 
in the correlated ground state. As a result, the collapse of the RPA 
(QRPA) at a critical value of the interaction parameter is avoided. 
However, it 
has been soon realized that the renormalized QRPA violates the Ikeda 
sum 
rule~\cite{Civi}. 
Several approaches were proposed recently 
to resolve this problem~\cite{ERRPA,res}.

The goal of this paper is to derive a modified HFB (MHFB) theory at 
finite 
temperature, which conserves the unitarity relation of the 
particle-density matrix. It will be shown that this can be 
achieved by using a modified quasiparticle-density matrix, which 
takes into account the statistical fluctuation of quasiparticle 
number 
microscopically. This modified quasiparticle-density matrix can be 
alternatively obtained by applying the secondary Bogoliubov 
transformation 
in Refs. \cite{MRPA,MBCS} on the particle-density matrix at 
zero temperature. 
It will be demonstrated that 
the BCS limit of the MHFB equations yields 
the modified BCS (MBCS) equations, which have been obtained 
previously in Refs.~\cite{MRPA,MBCS}. It will also be 
proved that the modified QRPA~\cite{MRPA}, 
obtained by using the MBCS quasiparticles, 
conserves the Ikeda sum rule.

The paper is organized as follows. Sec. II summarizes the main 
features of the FT-HFB theory and its violation of the unitarity 
relation. The MHFB theory, which restores 
the unitarity relation, is derived in Sec. III.  The MBCS 
equations are derived as the limit of the MHFB ones in the same 
section. The 
restoration of the Ikeda sum rule within the modified QRPA is shown 
in the 
Appendix.
The theory is illustrated in Sec. IV
by numerical calculations of the pairing gap and  
thermodynamic quantities such as the heat capacity, level-density 
parameter, and level density as functions of temperature for 
$^{120}$Sn. 
The same section also discusses in detail the applicability of the 
MBCS 
equations in numerical 
calculations using realistic single-particle energies at high 
temperature.
The paper is summarized in the last section, where 
conclusions are drawn.  
\section{Review of FT-HFB theory}
This section summarizes the main features of the FT-HFB theory, 
which has been derived by Goodman in Ref. \cite{Goodman1}. 
They are essential for 
deriving the MHFB theory at finite temperature in the present paper.

\subsection{HFB Hamiltonian}
The HFB theory is based on the self-consistent Hartree-Fock (HF) 
Hamiltonian with two-body interaction
\begin{equation}
H=\sum_{ij}{\cal T}_{ij}a_{i}^{\dagger}a_{j}+\frac{1}{4}\sum_{ijkl}
v_{ijkl}a_{i}^{\dagger}a_{j}^{\dagger}a_{l}a_{k}~,
\label{HHF}
\end{equation}
where $i,j,..$ denote the quantum numbers 
characterizing the single-particle 
orbitals, ${\cal T}_{ij}$ are the 
kinetic energies, and $v_{ijkl}$ are  
antisymmetrized matrix elements of the two-body interaction. 
The HFB theory approximates Hamiltonian (\ref{HHF}) by an 
independent-quasiparticle Hamiltonian $H_{\rm HFB}$
\begin{equation}
H-\mu\hat{N}\approx H_{\rm 
HFB}=E_{0}+\sum_{i}E_{i}\alpha_{i}^{\dagger}\alpha_{i}~,
\label{HHFB}
\end{equation}
where $\hat{N}$ is the particle-number operator, $\mu$ is the 
chemical 
potential, $E_{0}$ is the energy of the ground-state 
$|0\rangle$, which 
is defined as the vacuum of quasiparticles:
\begin{equation}
\alpha_{i}|0\rangle=0~,
\label{qpvacuum}
\end{equation}
and $E_{i}$ are quasiparticle energies. 
The quasiparticle creation
$\alpha_{i}^{\dagger}$ and destruction $\alpha_{i}$ operators are
obtained from the single-particle operators $a_{i}^{\dagger}$ and
$a_{i}$ by the Bogoliubov transformation,
whose matrix form is
\begin{equation}
	\left( \begin{array}{c}\alpha^{\dagger}\\
	\alpha\end{array}
	\right)=
	\left(\begin{array}{cc}
	U&V\\V^{*}&U^{*}
        \end{array}\right)
	\left(\begin{array}{c}
	a^{\dagger}\\ 
	a\end{array}\right)~
     \label{UV}
     \end{equation}
with the properties
\begin{equation}
UU^{\dagger}+VV^{\dagger}={\bf 1}~,\hspace{5mm} 
U{V}^{\rm T}+V{U}^{\rm T}=0~,
\label{UVproperties}
\end{equation}
where ${\bf 1}$ is the unit matrix, and the superscript $^{\rm T}$  
denotes the transposing operation. The two-body 
interaction term of Hamiltonian (\ref{HHF}), expressed in terms of 
quasiparticle operators using the transfromation (\ref{UV}), contains
also the terms 
$\sim\alpha^{\dagger}_{i}\alpha^{\dagger}_{j}
\alpha^{\dagger}_{k}\alpha^{\dagger}_{l}$, 
$\alpha^{\dagger}_{i}\alpha^{\dagger}_{j}
\alpha^{\dagger}_{k}\alpha_{l}$, 
$\alpha^{\dagger}_{i}\alpha^{\dagger}_{j}
\alpha_{l}\alpha_{k}$, and their hermitian conjugated parts. These 
terms are neglected in the HFB approximation. They play the role of 
residual interaction beyond the quasiparticle mean field.
The quasiparticle energies $E_{i}$ and matrices $U$ and $V$ are 
determined 
as the solutions of the HFB equations, which are usually derived by 
applying either 
the variational principle of Ritz or the Wick's 
theorem~\cite{RingSchuck}. 
\subsection{Thermodynamic and statistical quantities within 
FT-HFB theory}     
\label{thermo}
At finite temperature $T$ the condition for a system to be in 
thermal 
equilibrium requires the minimum of its grand potential $\Omega$
    \begin{equation}
    \Omega={\cal E}-TS-\mu N~,
    \label{Omega}
    \end{equation}
with the total energy ${\cal E}$, the entropy $S$, and particle 
number $N$, namely
\begin{equation}
    \delta\Omega=0~.
    \label{minimum}
    \end{equation}
    This variation defines the density operator ${\cal D}$ with
the trace equal to 1
 \begin{equation}
     {\rm Tr}{\cal D}=1~,\hspace{5mm} \delta\Omega/\delta{\cal D}=0~
     \label{dOmdD}
     \end{equation}
in the form
     \begin{equation}
	 {\cal D}=Z^{-1}{\rm e}^{-\beta(H-\mu\hat{N})},\hspace{5mm} 
	 Z={{\rm Tr}
	 [{\rm e}^{-\beta(H-\mu\hat{N})}]}~,\hspace{5mm} \beta=T^{-1}~,
	 \label{Dcal}
     \end{equation}
     where $Z$ is the grand partition function. The expectation value 
$\prec\hat{\cal O}\succ$ of 
     any operator $\hat{\cal O}$ is then given as the average in the 
grand canonical ensemble
     \begin{equation}
	 \prec\hat{\cal O}\succ = {\rm Tr}({\cal D}\hat{\cal O})~.
	 \label{average}
	 \end{equation}
	 This defines the total energy ${\cal E}$, entropy $S$, and 
	 particle number $N$ as 
     \begin{equation}
	 {\cal E}=\prec H\succ={\rm Tr}({\cal D}H)~,\hspace{3mm} 
	 S=-\prec{\cal D}{\rm ln}{\cal D}\succ=-{\rm Tr}({\cal D}{\rm 
ln}{\cal D})~,
	 \hspace{3mm} N=\prec{\hat{N}\succ}={\rm Tr}({\cal D}\hat{N})~.
	 \label{ESN}
	 \end{equation}	 
The FT-HFB theory replaces the unknown exact density operator 
${\cal D}$ in Eq. (\ref{Dcal}) 
with the approximated one, ${\cal D}_{\rm HFB}$, which is found in Ref. 
\cite{Goodman1} by substituting Eq. (\ref{HHFB}) in to Eq. 
(\ref{Dcal}) as
\begin{equation}
 {\cal D}_{\rm HFB}=
\prod_{i}[n_{i}\hat{\cal N}_{i}+(1-n_{i})
 (1-\hat{\cal N}_{i})]~,
 \label{DHFB}
 \end{equation}
where $\hat{\cal N}_{i}$ is the operator of quasiparticle number 
on the $i$-th orbital
     \begin{equation}
	 \hat{\cal N}_{i}=\alpha_{i}^{\dagger}\alpha_{i}~,
	 \label{Ncali}
	 \end{equation}
and $n_{i}$ is the quasiparticle occupation number. Within the FT-HFB 
theory $n_{i}$ is defined according to Eq. (\ref{average}) as
\begin{equation}
    n_{i}=\langle\hat{\cal N}_{i}\rangle= \frac{1}{{\rm e}^{\beta E_{i}}+1}~,
     \label{ni}
     \end{equation}
where the symbol $\langle\ldots\rangle$ denotes the 
average similar to (\ref{average}), but in which the approximated density operator 
${\cal D}_{\rm HFB}$ (\ref{DHFB}) replaces the exact one, i.e.
\begin{equation}
    \langle\hat{\cal O}\rangle={\rm Tr}({\cal D}_{\rm HFB}\hat{\cal O})~.
    \label{averageHFB}
    \end{equation}
That the quasiparticle occupation number $n_{i}$ at finite 
temperature is 
given by the Fermi-Dirac ditsribution as in 
Eq. (\ref{ni}) within the framework of the independent-quasiparticle 
approximation (\ref{HHFB}) has been also proved
a long time ago by Zubarev using the double-time Green function 
method~\cite{Zubarev} (See also the Appendix A of Ref. \cite{MBCS}).
The quasiparticle energy $E_{i}$ in Eq. (\ref{ni}) is found by solving 
the FT-HFB 
equations summarized in the next section. 
\subsection{FT-HFB equations}  
The generalized particle-density matrix $R$ is related to the 
generalized quasiparticle-density matrix $Q$ through the Bogoliubov 
transformation 
(\ref{UV}) as
\begin{equation}
R={\cal U}^{\dagger}Q{\cal U}~,
\label{RUQU}
\end{equation}
where
\begin{equation}
    R=\left(\begin{array}{cc}
	{\rho}&{\tau}\\-{\tau}^{*}&{1}-\rho^{*}
        \end{array}\right)~,\hspace{5mm} 
    Q=\left(\begin{array}{cc}
	{q}&{t}\\-{t}^{*}&{1}-{q}^{*}
        \end{array}\right)
	=\left(\begin{array}{cc}
	n&{0}\\{0}&{1}-n\end{array}\right)~,
	\label{RQ}
	\end{equation}
with
\begin{equation}
	{\cal U}=\left(\begin{array}{cc}
	U^{*}&V^{*}\\V&U
        \end{array}\right)~,\hspace{5mm} {\cal U}{\cal 
        U}^{\dagger}={\bf 1}~.
	\label{U}
	\end{equation}
The matrix elements of the single-particle matrix $\rho$ and particle 
pairing 
tensor $\tau$ within the FT-HFB approximation are evaluated as
\begin{equation}
    \rho_{ij}=\langle a_{j}^{\dagger}a_{i}\rangle~,\hspace{5mm} 
    \tau_{ij}=\langle a_{j}a_{i}\rangle~,
    \label{rot}
    \end{equation}
while those of the quasiparticle matrix $q$ are given in terms 
of the quasiparticle occupation number since
\begin{equation}    
q_{ij}=\langle\alpha_{j}^{\dagger}\alpha_{i}\rangle=\delta_{ij}n_{i}~,
    \hspace{5mm} t_{ij}=\langle\alpha_{j}\alpha_{i}\rangle=0~,
\label{q}
\end{equation}
which follow from the HFB approximation (\ref{HHFB}).
Using the inverse transformation 
of 
(\ref{UV}), the particle densities are obtained 
as~\cite{Goodman1}
\begin{equation}
\rho={U}^{\rm T}nU^{*}+V^{\dagger}(1-n)V~,\hspace{6mm} 
\tau={U}^{\rm T}nV^{*}+V^{\dagger}(1-n)U~.
\label{rhot}
\end{equation}
By minimizing the grand potential $\Omega$ according to Eq. 
(\ref{minimum}), 
Goodman has derived in Ref. \cite{Goodman1} the FT-HFB equations in 
the following form
\begin{equation}
 \left(\begin{array}{cc}
	{\cal H}&\Delta\\-\Delta^{*}&-{\cal H}^{*}
        \end{array}\right) \left(\begin{array}{c}
	U_{i}\\ 
	V_{i}\end{array}\right)=E_{i} \left(\begin{array}{c}
	U_{i}\\ 
	V_{i}\end{array}\right)~,
	\label{FTHFBeq}
	\end{equation}
where
\begin{equation}
    {\cal H}={\cal T}-\mu+\Gamma~,\hspace{6mm} 
    \Gamma_{ij}=\sum_{kl}v_{ikjl}\rho_{lk}~,\hspace{6mm}
    \Delta_{ij}=\frac{1}{2}\sum_{kl}v_{ijkl}\tau_{kl}~.
    \label{HGammaDelta}
    \end{equation}
(For the details of the derivation see Sec. 4 of Ref. 
\cite{Goodman1}).
The total energy ${\cal E}$, entropy $S$, and particle number $N$ 
from Eq. (\ref{ESN})
are now given within the FT-HFB theory as
\begin{equation}
    {\cal E}={\rm Tr}[({\cal T}+\frac{1}{2}\Gamma)\rho+\frac{1}{2}\Delta 
\tau^{\dagger}]~,
    \label{Ecal}
\end{equation}
\begin{equation}
    S=-\sum_{i}[n_{i}{\rm ln}n_{i}+(1-n_{i}){\rm ln}(1-n_{i})]~,
    \label{S}
\end{equation}
\begin{equation}
     N={\rm Tr}\rho~,
     \label{N}
\end{equation}
from which one can easily calculate the grand potential $\Omega$ 
(\ref{Omega}).

In the limit 
\begin{equation}
    v_{i~\widetilde{i}j~\widetilde{j}}=-G_{ij}~,
    \label{BCSlimit}
    \end{equation}
    where $|\widetilde{i}\rangle$ denotes the time-reversal state of 
    $|i\rangle$, 
    Eqs. (\ref{FTHFBeq}), (\ref{HGammaDelta}), and ({\ref{N}) yield 
the 
well-known FT-BCS equations. 
For spherical nuclei and with 
all the pairing matrix elements equal to $G_{ij}=G$, the FT-BCS 
equations have the form 
\begin{equation}
\Delta=G\sum_{j}\Omega_{j}u_{j}v_{j}(1-2n_{j})~,
\label{BCSgaps}
\end{equation}
\begin{equation}
    N=2\sum_{j}\Omega_{j}[(1-2n_{j})v_{j}^{2}+n_{j}]~,
    \label{BCSNs}
    \end{equation}
where $2\Omega_{j}=2j+1$ is the shell degeneracy. 
The quasiparticle energies $E_{j}$, and $u_{j}$ and $v_{j}$ 
coefficients are given as
\begin{equation}
E_{j}=\sqrt{(\epsilon_{j}-\mu)^{2}+\Delta^{2}}~,\hspace{5mm} 
u_{j}^{2}=\frac{1}{2}\bigg(1+\frac{\epsilon_{j}-\mu}{E_{j}}\bigg)~,\hspace{5mm} 
v_{j}^{2}=\frac{1}{2}\bigg(1-\frac{\epsilon_{j}-\mu}{E_{j}}\bigg)~.
\label{Euv}
\end{equation}    
\subsection{Violation of unitarity relation within FT-HFB theory}
At zero temperature ($T=$ 0) the quasiparticle occupation number 
vanishes: $n_{i}=$0, and the average (\ref{averageHFB}) reduces to the average in the quasiparticle 
vacuum (\ref{qpvacuum}). The quasiparticle-density 
matrix $Q$ (\ref{RQ}) becomes
\begin{equation}
    Q(T=0)\equiv Q_{0}=\left(\begin{array}{cc}
	0&0\\0&1
        \end{array}\right)~,\hspace{6mm} {\rm for}\hspace{2mm} {\rm which}\hspace{5mm} Q_{0}^{2}=Q_{0}~.
	\label{Q0}
	\end{equation}
Therefore, for the generalized particle-density matrix 
$R_{0}=R(T=0)$ the following unitarity relation
holds 
\begin{equation}
    R_{0}^{2}=R_{0}~,\hspace{5mm}
    \label{unitarity}
    \end{equation}
where
\begin{equation}
    R_{0}={\cal U}^{\dagger}Q_{0}{\cal U}~.
    \label{R0UQ0U}
    \end{equation}
However, the idempotent (\ref{unitarity}) no longer holds at $T\neq$ 0. 
Indeed, from Eqs. (\ref{RUQU}) and (\ref{RQ}) it follows that 
\begin{equation}
    R-R^{2}={\cal U}^{\dagger}(Q-Q^{2}){\cal U}~,
    \label{R-R2}
    \end{equation}
    which leads to
    \begin{equation}
	{\rm Tr}(R-R^{2})={\rm 
	Tr}(Q-Q^{2})=2\sum_{i}n_{i}(1-n_{i})\equiv 2(\delta{\cal N})^{2}\neq 
	0~,\hspace{5mm} (T\neq 0)~.
	\label{TrR-R2}
	\end{equation}
The quantity $\delta{\cal N}^{2}=\sum_{i}n_{i}(1-n_{i})$ in 
Eq. (\ref{TrR-R2}) is nothing but the 
quasiparticle-number fluctuation. This can be easily checked by 
calculating
\[
    \delta{\cal N}^{2}=\langle\hat{\cal N}^{2}
    \rangle-\langle\hat{\cal N}\rangle^{2}=\langle\sum_{i}\hat{\cal 
N}_{i}
    +\sum_{i\neq j}\hat{\cal N}_{i}
    \hat{\cal N}_{j}\rangle - \sum_{i}n_{i}^{2}-\sum_{i\neq 
    j}n_{i}n_{j}
    \]
  \begin{equation}  
    =\sum_{i}n_{i}(1-n_{i})  =\sum_{i}\delta{\cal N}_{i}^{2}~,
    \label{dN2}
    \end{equation}
    where
    \begin{equation}
	\delta{\cal N}_{i}^{2}=n_{i}(1-n_{i})~
	\label{dNi}
	\end{equation}
	is the fluctuation of quasiparticle number on the $i$-th orbital.
We've just seen that the violation of the unitarity relation 
(\ref{unitarity}) for the 
generalized single-particle density matrix $R$  
occurs at $T\neq$ 0 due to the fact that the HFB approximation (\ref{HHFB})
and the density operator ${\cal D}_{\rm HFB}$ (\ref{DHFB}) exclude 
the quasiparticle-number 
fluctuation (\ref{dN2}) from the quasiparticle-density matrix 
(\ref{RQ})~\cite{Goodman2}. Therefore, in order to restore the idempotent 
of type (\ref{unitarity}) at $T\neq$ 0 a new approximation should be found such that it includes 
the quasiparticle-number fluctuation [Eqs. (\ref{dN2}) and 
(\ref{dNi})] in the 
quasiparticle-density matrix. 
\section{Modified HFB (MHFB) theory at finite temperature} 
By including the quasiparticle-number fluctuation 
(\ref{dN2}), a part of the higher-order
terms $\sim\alpha_{i}^{\dagger}\alpha_{j}^{\dagger}\alpha_{k}\alpha_{l}$, 
neglected as the residual interaction beyond the FT-HFB quasiparticle mean field, will 
be taken into account. As the result the mean field of usual 
quasiparticles itself will be modified. This leads to 
the new quasiparticle energy $\bar{E}_{i}$ and chemical 
potential $\bar{\mu}$, which will be found as the solution of the 
modified HFB (MHFB) equations to be derived in this section. 
\subsection{Restoration of unitarity relation}
Let us consider, instead of the FT-HFB density operator ${\cal D}_{\rm 
HFB}$ (\ref{DHFB}), an improved approximation, $\bar{\cal D}$, to the density 
operator ${\cal D}$. This approximated density operator $\bar{\cal D}$ 
should satisfy two following requirements:

(a1) The average 
\begin{equation}
\langle\langle\hat{\cal O}\rangle\rangle\equiv{\rm Tr}(\bar{\cal D}\hat{\cal O}),
\label{averagebar}
\end{equation}
in which $\bar{\cal D}$ is used in place of ${\cal D}$ (or ${\cal D}_{\rm 
HFB}$), yields 
\begin{equation}
    \bar{R}={\cal U}^{\dagger}\bar{Q}{\cal U}~
    \label{RBARUQBARU}
    \end{equation}
for the 
Bogoliubov transformation ${\cal U}$ (\ref{U}), where one has the 
modified matrices
\begin{equation}
    \bar{R}=\left(\begin{array}{cc}
	\bar{\rho}&\bar{\tau}\\-\bar{\tau}^{*}&{1}-\bar\rho^{*}
        \end{array}\right)~,\hspace{5mm} 
    \bar{Q}=\left(\begin{array}{cc}
	\bar{q}&{t}\\-\bar{t}^{*}&{1}-\bar{q}^{*}
        \end{array}\right)~,
	\label{RQbar}
	\end{equation}
with	
\begin{equation}
    \bar{\rho}_{ij}=\langle\langle a_{j}^{\dagger}a_{i}\rangle\rangle~,\hspace{5mm} 
    \bar{\tau}_{ij}=\langle\langle a_{j}a_{i}\rangle\rangle~,
    \label{rotbar}
    \end{equation}
\begin{equation}    
\bar{q}_{ij}=\langle\langle\alpha_{j}^{\dagger}\alpha_{i}\rangle\rangle=
\delta_{ij}\bar{n}_{i}~,
    \hspace{5mm} 
 \bar{t}_{ij}=\langle\langle\alpha_{j}\alpha_{i}\rangle\rangle={\bf\Lambda}_{ij}~
\label{qbar}
\end{equation}
instead of matrices $R$ and $Q$ in Eqs. (\ref{RQ}), (\ref{rot}), and 
(\ref{q}).
The non-zero values of $\bar{t}_{ij}$ in 
Eq. (\ref{qbar}) are caused by the quasiparticle correlations in the 
thermal equilibrium, which are now included in the average 
$\langle\langle\ldots\rangle\rangle$ using the density operator 
$\bar{\cal D}$.

(a2) The modified quasiparticle-density matrix $\bar{Q}$ satisfies the 
unitarity relation   
\begin{equation}
(\bar{Q})^{2}=\bar{Q}~.
\label{Qbar2}
\end{equation}
The solution of Eq. (\ref{Qbar2}) immediately yields 
the matrix $\bf\Lambda$ in the canonical form
\begin{equation}
{\bf\Lambda}=\sqrt{\bar{n}(1-\bar{n})}\equiv
\left(\begin{array}{cccccccccccc}
	&&&&&&&&&&0&-\Lambda_{1}\\
	&&&&&&&&&&\Lambda_{1}&0\\
	&&&&&&&&0&-\Lambda_{2}&&\\
        &&&&&&&&\Lambda_{2}&0&&\\
        &&&&&&.&&&&\\
        &&&.&&&&&&&\end{array}\right)~,\hspace{5mm} 
        \Lambda_{i}=\sqrt{\bar{n}_{i}(1-\bar{n}_{i})}~.
\label{x}
\end{equation}
Comparing this result with Eqs. (\ref{dN2}) and (\ref{dNi}), it is clear 
that tensor ${\bf\Lambda}$ consists of the quasiparticle-number 
fluctuation $\delta\bar{\cal N}_{i}=\sqrt{\bar{n}_{i}(1-\bar{n}_{i})}$.
From Eq. (\ref{RBARUQBARU}) 
it is easy to see that the unitarity relation holds for the 
modified generalized single-particle density matrix $\bar{R}$ since
\begin{equation}
    \bar{R}-\bar{R}^{2}={\cal U}^{\dagger}(\bar{Q}-\bar{Q}^{2}){\cal U}=0~
    \label{R-R2b}
    \end{equation}
due to Eq. (\ref{Qbar2}) and the unitary matrix ${\cal U}$. 

Let us define the modified-quasiparticle operators 
$\bar{\alpha}^{\dagger}_{i}$ and $\bar{\alpha}_{i}$, which behave in
the average (\ref{averagebar}) exactly as the usual quasiparticle 
operators 
$\alpha_{i}^{\dagger}$ and $\alpha_{i}$ do in the quasiparticle 
ground state, namely
\begin{equation}            
    \langle\langle\bar{\alpha}^{\dagger}_{i}\bar{\alpha}_{k}\rangle\rangle=
    \langle\langle\bar{\alpha}^{\dagger}_{i}\bar{\alpha}_{k}^{\dagger}\rangle\rangle=
    \langle\langle\bar{\alpha}_{k}\bar{\alpha}_{i}\rangle\rangle=0~.
    \label{avealphabar}
    \end{equation}
In the same way as for the usual Bogoliubov transformation (\ref{UV}),
we search for a transformation between these modified-quasiparticle 
operators ($\bar{\alpha}^{\dagger}_{i}$, $\bar{\alpha}_{i}$) and the 
usual quasiparticle ones  (${\alpha}^{\dagger}_{i}$, ${\alpha}_{i}$)
in the following form 
\begin{equation}
	\left( \begin{array}{c}\bar{\alpha}^{\dagger}\\
	\bar{\alpha}\end{array}
	\right)=
	\left(\begin{array}{cc}
	w&z\\z^{*}&w^{*}
        \end{array}\right)
	\left(\begin{array}{c}
	\alpha^{\dagger}\\ 
	\alpha\end{array}\right)~,
     \label{Bogo2}
     \end{equation}
with the unitary property similar to Eq. (\ref{UVproperties}) 
for $U$ and $V$ matrices
\begin{equation}
    ww^{\dagger}+zz^{\dagger}={\bf 1}~.
    \label{ww+zz}
    \end{equation}
Using the inverse transformation of (\ref{Bogo2}) and the 
requirement (\ref{avealphabar}), we obtain
\begin{equation}
    \bar{n}_{i}=\langle\langle\alpha^{\dagger}_{i}\alpha_{i}\rangle\rangle= 
    \sum_{k}z_{ik}z_{ik}^{*}~.
    \label{nz}
    \end{equation}
From this equation and the unitary condition (\ref{ww+zz}), it 
follows that $zz^{\dagger}=\bar{n}$ and $ww^{\dagger}={\bf 1}-\bar{n}$. Since 
${\bf 1}-\bar{n}$ and $\bar{n}$ are real diagonal matrices, 
the canonical form of matrices $w$ and $z$ is found as
\begin{subequations}    
    \label{wzmatrices}  
\begin{equation}
    w=\sqrt{1-\bar{n}}\equiv\left(\begin{array}{ccccccccccc}
	w_{1}&0&&&&&&&&&\\
	0&w_{1}&&&&&&&&&\\
	&&w_{2}&0&&&&&&&\\
        &&0&w_{2}&&&&&&&\\
        &&&&.&&&&&\\
        &&&&&&&.&&\end{array}\right)~,\hspace{4mm} 
	z=\sqrt{\bar{n}}\equiv\left(\begin{array}{ccccccccccc}
	&&&&&&&&&0&-z_{1}\\
	&&&&&&&&&z_{1}&0\\
	&&&&&&&0&-z_{2}&&\\
        &&&&&&&z_{2}&0&&\\
        &&&&&&.&&&&\\
        &&&&.&&&&&&\end{array}\right)~, 
    \end{equation}
    where
\begin{equation}
        w_{i}=\sqrt{1-\bar{n}_{i}}~~~~~,\hspace{40mm} 
        z_{i}=\sqrt{\bar{n}_{i}}~.
\end{equation}
\end{subequations}
We see now that, just like Eq. (\ref{UV}), which is the 
generalized form of the Bogoliubov transformation for the BCS case, 
Eq. (\ref{Bogo2}) with matrices $w$ and $z$ uniquely
defined in Eq. (\ref{wzmatrices}) is the genearlized form of the 
secondary Bogoliubov transformation used in Refs. \cite{MRPA,MBCS}. 
It expresses a simple relationship between the 
modified-quasiparticle operators 
($\bar{\alpha}^{\dagger}_{i}$, $\bar{\alpha}_{i}$)
and the usual-quasiparticle operators
(${\alpha}^{\dagger}_{i}$, ${\alpha}_{i}$) in the 
same fashion as that between the latter and the single-particle 
operators in the Bogoliubov transformation (\ref{UV}).

We now show that we can obtain the idempotent 
$\bar{R}^{2}=\bar{R}$ (\ref{R-R2b}) by applying 
the secondary Bogoliubov 
transformation (\ref{Bogo2}), which automatically leads to Eq. 
(\ref{Qbar2}).  
Indeed, using the inverse transformation of (\ref{Bogo2}) with matrices 
$w$ and $z$ given in 
Eq. (\ref{wzmatrices}), we found that the 
modified quasiparticle-density matrix $\bar{Q}$ can be obtained as
\begin{equation}
    {\cal W}^{\dagger}\bar{Q}_{0}{\cal W}=\left(\begin{array}{cc}
	\bar{n}&[\sqrt{\bar{n}(1-\bar{n})}]
	^{\dagger}\\\sqrt{\bar{n}(1-\bar{n})}&1-\bar{n}\end{array}\right)
	    \equiv\bar{Q}~,
	\label{WQ0W}
\end{equation}
where
\begin{equation}
    {\cal W}=\left(\begin{array}{cc}	
(\sqrt{1-\bar{n}})^{*}&(\sqrt{\bar{n}})^{*}\\\sqrt{\bar{n}}&\sqrt{1-\bar{n}}
\end{array}\right)~,	
\hspace{5mm} {\cal W}{\cal W}^{\dagger}={\bf 1}~,
\label{W}
\end{equation}
and
\begin{equation}
    \bar{Q}_{0}=\left(\begin{array}{cc}	
\langle\langle\bar{\alpha}^{\dagger}\bar{\alpha}\rangle\rangle
&\langle\langle\bar{\alpha}\bar{\alpha}\rangle\rangle\\
\langle\langle\bar{\alpha}^{\dagger}\bar{\alpha}^{\dagger}
\rangle\rangle&1-\langle\langle\bar{\alpha}^{\dagger}\bar{\alpha}\rangle\rangle
\end{array}\right)=
\left(\begin{array}{cc}	
0&0\\0&1
\end{array}\right)~,\hspace{5mm} \bar{Q}_{0}^{2}=\bar{Q}_{0}~,
\label{Q0bar}
\end{equation}
due to Eq. (\ref{avealphabar}). This result shows another way
of deriving the
modified quasiparticle-density matrix $\bar{Q}$ (\ref{RQbar})
from the density matrix $\bar{Q}_{0}$ of the modified quasiparticles
($\bar{\alpha}^{\dagger}_{i}$, $\bar{\alpha}_{i}$). This matrix 
$\bar{Q}_{0}$ is identical to
the zero-temperature quasiparticle-density matrix $Q_{0}$ (\ref{Q0}).
Substituting this result into the right-hand side (r.h.s) of 
Eq. (\ref{RBARUQBARU}), we obtain
\begin{equation}
\bar{R}=\bar{\cal U}^{\dagger}\bar{Q}_{0}\bar{\cal U}~,
\label{RUBARQ0UBAR}
\end{equation}
where
\begin{equation}
\bar{\cal U}={\cal W}{\cal U}=
 \left(\begin{array}{cc}	
\bar{U}^{*}&\bar{V}^{*}\\\bar{V}&\bar{U}\end{array}\right)
=
 \left(\begin{array}{cc}	
(\sqrt{1-\bar{n}})^{*}U^{*}+(\sqrt{\bar{n}})^{*}V
&~~~~(\sqrt{1-\bar{n}})^{*}V^{*}+(\sqrt{\bar{n}})^{*}U\\
\sqrt{1-\bar{n}}V+\sqrt{\bar{n}}U^{*}&~~~~\sqrt{1-\bar{n}}U+\sqrt{\bar{n}}V^{*}\end{array}\right)~.
\label{Ubar}
 \end{equation}
This equation is the generalized form of the modified Bogoliubov 
coefficients $\bar{u}_{j}$ and $\bar{v}_{j}$ given in Eq. (6) of 
Ref. \cite{MRPA} or Eq. (38) of Ref. \cite{MBCS}.  From Eqs. 
(\ref{U}), (\ref{W}), and (\ref{Ubar}), it follows that 
$\bar{\cal U}\bar{\cal U}^{\dagger}={\bf 1}$, i.e.  
transformation (\ref{RUBARQ0UBAR}) is unitary.
Therefore, from the idempotent (\ref{Q0bar}) it follows 
that $\bar{R}^{2}=\bar{R}$. 
We have just shown 
that the secondary Bogoliubov transformation (\ref{Bogo2}) allows us 
to take into account the fluctuation of quasiparticle number and restore 
the unitarity relation of the generalized particle-density matrix
~\footnote{An alternative approach to the unitarity problem 
was propposed in Ref. \cite{Tanabe2} 
making use of the 
thermo field dynamics~\cite{TFD}.}. 
In this sense, the approximation discussed in the present section is a step
beyond the thermal mean field of usual quasiparticles. As the result, 
the thermal quasiparticle mean field, which was defined within the 
FT-HFB approximation, is modified due to thermal 
quasiparticle-number fluctuation.
\footnote
{An exact theory on quasiparticle excitations at $T=$ 0 should 
define the vacuum and quasiparticles in terms of exact eigenstates of the 
many-body system~\cite{RingSchuck}. 
But in this case, a simple 
mathematical relationship between the exact quasiparticles and 
the usual particles of the system no longer exists. 
The advantage of the Bogoliubov-type quasiparticles 
is the linear relationship between them and 
the usual particles.
However, the corresponding vaccum and 
single-quasiparticle state are now only approximations of the exact 
eigen functions of the many-body Hamiltonian. 
Similarly, at $T\neq$ 0, when the average over the individual 
compound systems is replaced by that over the grand canonical 
ensemble (\ref{average}), the density operators 
${\cal D}_{\rm HFB}$ (\ref{DHFB}) and $\bar{\cal D}$ 
(\ref{averagebar}) are different approximations of the exact density operator 
${\cal D}$ (\ref{Dcal}).}

\subsection{MHFB equations at finite temperature}
With all the thermal degrees of freedom 
now included in $\bar{\cal U}$, Eq. (\ref{RUBARQ0UBAR}) formally 
looks 
the same as the usual HFB approximation at $T=0$ (\ref{R0UQ0U}), which connects 
$R_{0}$ 
to $Q_{0}$. 
Applying the Wick's theorem for the ensemble 
average~\cite{RingSchuck}, one obtains 
the expressions for the modified 
total energy $\bar{\cal E}$
\begin{equation}
\bar{\cal E}={\rm 
Tr}[({\cal T}+\frac{1}{2}\bar{\Gamma})\bar{\rho}+\frac{1}{2}
\bar{\Delta}\bar{\tau}^{\dagger}]~,
\label{Ecalbar}
\end{equation}
where
\begin{equation}
\bar{\Gamma}_{ij}=\sum_{kl}v_{ikjl}\bar{\rho}_{lk}~,
\label{Gammabar}
\end{equation}
\begin{equation}
\bar{\Delta}_{ij}=\frac{1}{2}\sum_{kl}v_{ijkl}\bar{\tau}_{kl}~.
\label{Deltabar}
\end{equation}
From Eq. (\ref{RUBARQ0UBAR}) we obtain the modified single-particle 
denity matrix $\bar{\rho}$ and modified
particle-pairing tensor $\bar{\tau}$ in the following form
\begin{equation}
\bar{\rho}={U}^{\rm T}\bar{n}U^{*}+V^{\dagger}(1-\bar{n})V+
{U}^{\rm 
T}\bigg[\sqrt{\bar{n}(1-\bar{n})}\bigg]^{\dagger}V+V^{\dagger}\sqrt{\bar{n}(1-
\bar{n})}U^{*}~,
\label{rhobar}
\end{equation}
\begin{equation}
\bar{\tau}={U}^{\rm T}\bar{n}V^{*}+V^{\dagger}(1-\bar{n})U
+{U}^{\rm 
T}\bigg[\sqrt{\bar{n}(1-\bar{n})}\bigg]^{\dagger}U+V^{\dagger}\sqrt{\bar{n}(1-\bar{n})}
V^{*}~.
\label{taubar}
\end{equation}
As compared to
Eq. (\ref{rhot}) within the FT-HFB 
approximation, Eqs. (\ref{rhobar}) and (\ref{taubar}) 
contain the last two terms  
$\sim[\sqrt{\bar{n}(1-\bar{n})}]^{\dagger}$ and $\sim\sqrt{\bar{n}(1-\bar{n})}$, 
which arise due to quasiparticle-number 
fluctuation. Also the quasiparticle occupation number is now 
$\bar{n}$ [See Eq. (\ref{qbar})]
instead of $n$ (\ref{ni}).

We derive the MHFB equations following the same 
variational procedure, which was used to derive the FT-HFB 
equations in Sec. 4 of Ref. \cite{Goodman1}. According it, we minimize 
the grand potential $\delta{\bar{\Omega}}=$ 0 by varying $U$, $V$, and 
$\bar{n}$, where 
\begin{equation}
\bar{\Omega}=\bar{\cal 
E}-T\bar{S}-\bar{\mu}N~.
\label{Omegabar}
\end{equation}
Due to Eq. (\ref{UVproperties}), the variations $\delta U$ and $\delta 
V$ are not independent. They are found by using an infinitesimal 
unitary tranformation of (\ref{UV}). 
The obtained infinitesimal variations $U'=U+\delta U$ and $V'=V+\delta 
V$ together with $\bar{n}'=\bar{n}+\delta\bar{n}$ 
are then used in  Eqs. (\ref{rhobar}) and (\ref{taubar}) to 
obtain $\bar{\rho}'=\bar{\rho}+\delta\bar{\rho}$ and 
$\bar{\tau}'=\bar{\tau}+\delta{\tau}$. Substituting them into
Eq. (\ref{Omegabar}) one obtains 
$\bar{\Omega}'=\bar{\Omega}+\delta\bar\Omega$, where 
$\delta\bar{\Omega}$ is expressed in terms of $\delta\bar{\rho}$, 
$\delta\bar{\tau}$, and $\delta\bar{n}$ as independent variations.
By requiring the coefficients of $\delta\bar{\rho}$ and 
$\delta\bar{\tau}$ vanish and following the rest of the derivation as 
for the zero-temperature case, we finally obtain the MHFB 
equations, which formally look like the FT-HFB ones 
(\ref{FTHFBeq})
\begin{equation}
 \left(\begin{array}{cc}
	\bar{\cal H}&\bar{\Delta}\\-\bar{\Delta}^{*}&-\bar{\cal H}^{*}
        \end{array}\right) \left(\begin{array}{c}
	U_{i}\\ 
	V_{i}\end{array}\right)=\bar{E}_{i} \left(\begin{array}{c}
	U_{i}\\ 
	V_{i}\end{array}\right)~,
	\label{MHFBeq}
	\end{equation}
where, however
\begin{equation}
    \bar{\cal H}={\cal T}-\bar{\mu}+\bar{\Gamma}~
    \label{Hbar}
    \end{equation}
with $\bar{\Gamma}$ and $\bar{\Delta}$ given by Eqs. 
 (\ref{Gammabar}) and (\ref{Deltabar}), respectively.
 The equation for particle number $N$ within the MHFB theory is
 \begin{equation}
     N={\rm Tr}\bar{\rho}~.
     \label{NMHFB}
     \end{equation}
By solving Eq. (\ref{MHFBeq}), one obtains the modified quasiparticle energy 
$\bar{E}_{i}$, which is different from $E_{i}$ 
in Eqs (\ref{FTHFBeq}) and/or (\ref{Euv}) due to 
the change of the HF and pairing potentials. Hence, 
the MHFB quasiparticle Hamiltonian $H_{\rm MHFB}$ can be 
written as
\begin{equation}
    H-\bar{\mu}\hat{N}\approx H_{\rm MHFB}=\bar{E}_{0}+\sum_{i}\bar{E}_{i}\hat{\cal N}_{i}~,
    \label{HMHFB}
    \end{equation}
    instead of (\ref{HHFB}).
This implies that
the approximated density operator 
$\bar{D}$ (\ref{averagebar}) within the MHFB theory 
can be represented in the form similar 
to (\ref{DHFB}), namely
\begin{equation}
 \bar{\cal D}\equiv{\cal D}_{\rm MHFB}=
 \prod_{i}[\bar{n}_{i}\hat{\cal N}_{i}+(1-\bar{n}_{i})
 (1-\hat{\cal N}_{i})].~
 \label{DMHFB}
 \end{equation}  
From here it follows that the formal expression for the modified entropy 
$\bar{S}$ is the same as that given in Eq. (\ref{S}), i.e.
\begin{equation}
    \bar{S}=-\sum_{i}[\bar{n}_{i}{\rm ln}\bar{n}_{i}+(1-\bar{n}_{i}){\rm 
    ln}(1-\bar{n}_{i})]~,
    \label{barS}
\end{equation}
Using the thermodynamic definition of temperature in terms of entropy 
$T=\delta{\bar{S}}/\delta{\bar{\cal E}}$ and carrying out the 
variation over $\delta\bar{n}_{i}$, we find
\begin{equation}
\frac{\delta\bar{\cal E}}{\delta\bar{n}_{i}}\equiv
\bar{E}_{i}=T\frac{\delta\bar{S}}{\delta\bar{n}_{i}}=
T{\rm ln}\bigg(\frac{1-\bar{n}_{i}}{\bar{n}_{i}}\bigg)~.
\label{varE}
\end{equation}
Inverting Eq. (\ref{varE}), we obtain
\begin{equation}
    \bar{n}_{i}=\frac{1}{{\rm e}^{\beta\bar{E}_{i}+1}}~.
    \label{nbar}
    \end{equation}
This result shows that the functional dependence of 
quasiparticle occupation number 
$\bar{n}_{i}$ on quasiparticle energy and temperature within the MHFB theory 
is also given by the 
Fermi-Dirac distribution of noninteracting quasiparticles but with 
the modified energies $\bar{E}_{i}$ defined by the MHFB equations 
(\ref{MHFBeq})~\footnote{Note that, 
there remains the residual interaction, even due to pairing alone, beyond 
the MHFB quasiparticle mean field. At $T=$ 0 this can be treated as 
ground state correlations within the 
renormalized and/or modified QRPA~\cite{MRPA,RRPA}. 
As the result $\bar{n}_{i}$ deviates from the Fermi-Dirac 
distribution, especially if different 
multipolarities of the two-body residual interaction 
are taken into account. However, for the monopole 
pairing interaction alone as considered in this paper, 
such deviation is negligible [See Appendix B of Ref. \cite{MBCS}].
At $T\neq$ 0 the quasiparticle-number 
fluctuation beyond the quasiparticle mean field 
leads to the entropy effect within the renormalized RPA, 
which was studied in Ref. \cite{Entro}.}. 
Therefore in the rest of the paper we will omit the bar over 
$\bar{n}_{i}$ and use 
the same Eq. (\ref{ni}) with ${E}_{i}$ replaced
with $\bar{E}_{i}$ for the MHFB equations.  
 \subsection{Modified BCS (MBCS) theory at finite temperature}
 \subsubsection{MBCS equations}
 The MBCS equations at finite temperature 
 have been derived previously in Refs. 
 \cite{MRPA,MBCS} using the  
 secondary Bogoliubov transformation (\ref{Bogo2}) for the BCS case. 
 We will show below that
 these MBCS equations emerge as the 
 limit of the MHFB equations 
 derived in the preceding section.
 
 In the BCS limit (\ref{BCSlimit}) with equal pairing matrix elements 
 $G_{ij}=G$, 
 neglecting the contribution of
 $G$ to the HF potential so that $\bar{\Gamma}=$ 0, the HF Hamiltonian
 becomes
 \begin{equation}
     \bar{\cal H}_{ij}=(\epsilon_{i}-\bar{\mu})\delta_{ij}~.
     \label{HHFBCS}
     \end{equation}
 The pairing potential (\ref{Deltabar}) takes now the simple form 
 \begin{equation}
      \bar{\Delta}=-G\sum_{k>0}\bar{\tau}_{k\widetilde{k}}~.
     \label{DeltaMBCS}
     \end{equation}
 The Bogoliubov transformation (\ref{UV}) for spherical nuclei 
reduces to
 \[
 {\alpha}_{jm}^{\dagger}=u_{j}a_{jm}^{\dagger}+v_{j}(-)^{j+m}a_{j-m}~,
 \]\begin{equation}
 (-)^{j+m}{\alpha}_{j-m}=u_{j}(-)^{j+m}
 a_{j-m}-v_{j}a_{jm}^{\dagger}~,
 \label{uv}
 \end{equation}
 while the secondary Bogoliubov transformation (\ref{Bogo2}) 
 becomes~\cite{MRPA,MBCS}
 \[
\bar{\alpha}_{jm}^{\dagger}=\sqrt{1-n_{j}}\alpha_{jm}^{\dagger}-\sqrt{n_{j}}
(-)^{j+m}{\alpha}_{j-m}~,
 \]
 \begin{equation}
 (-)^{j+m}\bar{\alpha}_{j-m}=\sqrt{1-n_{j}}(-)^{j+m}
 {\alpha}_{j-m}+\sqrt{n_{j}}{\alpha}_{jm}^{\dagger}~.
 \label{w}
 \end{equation}
 The $U$, $V$, $1-n$, $n$, and $\sqrt{n(1-n)}$ matrices are 
 now block diagonal in each two-dimensional subspace spanned by the 
 quasiparticle state $|j\rangle$ 
 and its time-reversal partner 
$|\widetilde{j}\rangle=(-)^{j+m}|j-m\rangle$
 \begin{equation}
  U= \left(\begin{array}{cc}
	u_{j}&0\\0&u_{j}
        \end{array}\right)~,\hspace{5mm}  V= \left(\begin{array}{cc}
	0&v_{j}\\-v_{j}&0
        \end{array}\right)~,
	\label{UiVi}
	\end{equation}
\[
  1-n= \left(\begin{array}{cc}
	1-n_{j}&0\\0&1-n_{j}
        \end{array}\right)~,\hspace{5mm}  n= \left(\begin{array}{cc}
	n_{j}&0\\0&n_{j}
        \end{array}\right)~,\]
\begin{equation} \sqrt{n(1-n)}= 
	\left(\begin{array}{cc}
	0&-\sqrt{n_{j}(1-n_{j})}\\\sqrt{n_{j}(1-n_{j})}&0
        \end{array}\right)~,
	\label{nmatrices}
	\end{equation}
Substituting these matrices into the r.h.s of Eqs. (\ref{rhobar}) 
and (\ref{taubar}), 
we find
\begin{equation}
\bar{\rho}_{j~\widetilde{j}}=(1-2n_{j})v_{j}^{2}+n_{j}-2\sqrt{n_{j}(1-n_{j})}
u_{j}v_{j}~,
\label{rhoMBCS}
\end{equation}
\begin{equation}
\bar{\tau}_{j~\widetilde{j}}=-(1-2n_{j})u_{j}v_{j}+\sqrt{n_{j}(1-n_{j})}(u_{j}^{2}-
v_{j}^{2})~.
\label{tauMBCS}
\end{equation}
Substituting now Eqs. (\ref{tauMBCS}) and (\ref{rhoMBCS}) into the 
r.h.s of Eqs. 
(\ref{DeltaMBCS}) and (\ref{NMHFB}), respectively, we obtain the MBCS 
equations for spherical nuclei in the following form:
\begin{equation}
 \bar{\Delta}=G\sum_{j}\Omega_{j} 
 [(1-2n_{j})u_{j}v_{j}-\sqrt{n_{j}(1-n_{j})}(u_{j}^{2}-v_{j}^{2})]~,
 \label{MBCSgap}
 \end{equation}
 \begin{equation}    
N=2\sum_{j}\Omega_{j}[(1-2n_{j})v_{j}^{2}+n_{j}-2\sqrt{n_{j}(1-n_{j})}
    u_{j}v_{j}]~.
 \label{MBCSN}
 \end{equation}
Equations (\ref{MBCSgap}) and
(\ref{MBCSN}) are exactly the same as 
the MBCS Eqs. (23) and (24) in Ref. \cite{MRPA} or Eqs. (39) and (40) 
in 
Ref. \cite{MBCS}. We've just shown that the MBCS equations in Refs. 
\cite{MRPA,MBCS} emerge as the natural limit of the MHFB equations 
at finite-temperature.

For convenience in further discussions we rewrite the MBCS gap in 
Eq. (\ref{MBCSgap}) as a sum of quantal $\Delta_{\rm Q}$ 
and thermal-fluctuation $\delta\Delta$ parts as
\begin{equation}
 \bar{\Delta}=\Delta_{\rm Q}+\delta\Delta~,
 \label{MBCSgap1}
 \end{equation}
where the quantal gap $\Delta_{\rm Q}$ is
 \begin{equation}
\Delta_{\rm Q}=\sum_{j}(\Delta_{\rm Q})_{j}~,\hspace{5mm} 
     (\Delta_{\rm Q})_{j}=G\Omega_{j}u_{j}v_{j}(1-2n_{j})~.
     \label{DQ}
     \end{equation}
It is called quantal since it is caused by quantal effects 
starting from $T=$ 0, where it is 
equal to the BCS gap, and decreases as $T$ increases 
because the Pauli blocking becomes weaker.
The thermal-fluctuation gap $\delta\Delta$, referred to hereafter as 
thermal gap, is given as
\begin{equation}
  \delta\Delta=\sum_{j}\delta\Delta_{j}~,\hspace{5mm} 
  \delta\Delta_{j}=G\Omega_{j}(v_{j}^{2}-u_{j}^{2})\delta{\cal 
N}_{j}~,
\label{dD}
\end{equation}
and arises due to the thermal quasiparticle-number fluctuation 
$\delta{\cal N}_{j}$ at 
$T\neq$ 0.
Therefore, 
comparing the FT-BCS equations (\ref{BCSgaps}) and (\ref{BCSNs}) 
with the 
MBCS ones, (\ref{MBCSgap}) and (\ref{MBCSN}), we see that 
the latter explicitly include the effect of 
quasiparticle-number fluctuation $\sim\delta{\cal N}_{j}$ (\ref{dNi})
in the last terms at their r.h.s, which are the thermal gap 
(\ref{dD}) 
in Eq. (\ref{MBCSgap}) and the thermal-fluctuation of particle number 
$\delta N=\sum_{j}\delta N_{j}=
-4\sum_{j}\Omega_{j}u_{j}v_{j}(\delta{\cal N}_{j})$ 
in Eq. (\ref{MBCSN}).
These terms are ignored within 
the FT-BCS theory. Hence Eqs. (\ref{MBCSgap}) and 
(\ref{MBCSN}) show for the first time how the effect of statistical 
fluctuations is included in the MBCS (MHFB) theory at finite 
temperature 
on a microscopic ground.
So far this effect was treated only within the framework
of the macroscopic Landau theory of phase transition~\cite{Moretto}. 
\subsubsection{Thermodynamics quantities}
The total energy $\bar{\cal E}$ is found as
\begin{equation}
\bar{\cal E}=2\sum_{j}\Omega_{j}\epsilon_{j}[(1-2n_{j})v_{j}^{2}+n_{j}
-2\sqrt{n_{j}(1-n_{j})}u_{j}v_{j}]-\frac{\bar{\Delta}^{2}}{G}~.
\label{Ebar}
\end{equation}
The heat capacity $C$ is calculated as the derivative of energy 
$\bar{\cal E}$ (\ref{Ebar}) with respect to temperature $T$
\begin{equation}
C=\frac{\partial\bar{\cal E}}{\partial T}~.
\label{C}
\end{equation}
The level-density parameter $a$ is defined by 
the Fermi-gas formula
\begin{equation}
a=\frac{E^{*}}{T^{2}}=\frac{\bar{\cal E}(T)-\bar{\cal E}(0)}{T^{2}}~,
\label{a-level}
\end{equation}
where $E^{*}\equiv\bar{\cal E}(T)-\bar{\cal E}(0)$ is the excitation 
energy of 
the system.
The quasiparticle entropy (\ref{barS}) is written 
for spherical nuclei as
\begin{equation}
\bar{S}=-2\sum_{j}\Omega_{j}[n_{j}{\rm ln}n_{j}+(1-n_{j}){\rm 
ln}(1-n_{j})]
=2\sum_{j}\Omega_{j}\bigg[\frac{\beta\bar{E}_{j}}
{{\rm e}^{\beta\bar{E}_{j}}+1}+{\rm ln}(1+{\rm 
e}^{-\beta\bar{E}_{j}})\bigg]~.
\label{Sbar}
\end{equation}
Using the MBCS equations (\ref{MBCSgap}) and (\ref{MBCSN}), Eqs. 
(\ref{Ebar}), and (\ref{Sbar}) together with the expressions for
$\bar{E}_{i}$, $u_{i}$, and $v_{i}$, which are the same as in 
Eq. (\ref{Euv}) (with $\bar{E}_{i}$ replacing $E_{i}$ and 
$\bar{\Delta}$ replacing $\Delta$), we found that the formal 
expression for the grand potential $\Phi$ is also the same 
as that given within the FT-BCS theory~\cite{Sano,Moretto2}, 
namely
\begin{equation}
\Phi\equiv-\beta\bar{\Omega}= 
-\beta\sum_{j}\Omega_{j}[\epsilon_{j}-\bar{\mu}-\bar{E}_{j}]
+2\sum_{j}\Omega_{j}{\rm ln}[1+{\rm e}^{-\beta\bar{E}_{j}}] 
-\beta\frac{\bar{\Delta}^{2}}{G}~.
\label{grandpotMBCS}
\end{equation}
The level density $\rho(N, Z)$ 
is calculated as the inverse Laplace transform of 
the grand partition function ${\rm e}^{\Phi}$. It is approximated 
as~\cite{Moretto2,Behkami}
\begin{equation}
\rho(N, Z)=\frac{{\rm 
e}^{\bf S}}{2\pi\sqrt{2\pi{\bf D}}}~,
\label{leveldensity}
\end{equation}
where ${\bf S}=\bar{S}_{N}+\bar{S}_{Z}$ is the total entropy of the 
system, ${\bf D}$ is the determinant of the second derivatives of the 
grand partition function taken at the 
saddle point. It is given as
\begin{equation}
{\bf D}=\frac{\partial^{2}{\Phi}}{\partial\alpha_{N}^{2}}{\bf 
D}_{Z}   
+\frac{\partial^{2}{\Phi}}{\partial\alpha_{Z}^{2}}{\bf D}_{N}~,
\hspace{5mm} 
{\bf D}_{i}=\left|\begin{array}{cc}
\frac{\partial^{2}{\Phi}_{i}}{\partial\alpha_{i}^{2}} & 
\frac{\partial^{2}{\Phi}_{i}}{\partial\alpha_{i}\partial\beta} 
\\ 
\frac{\partial^{2}{\Phi}_{i}}{\partial\alpha_{i}\partial\beta} 
& 
\frac{\partial^{2}{\Phi}_{i}}{\partial\beta^{2}}
\end{array}\right|~,\hspace{5mm} i=N, Z~,\hspace{5mm} 
\alpha_{i}={\beta}\bar{\mu}_{i}~.
\label{bdf}
\end{equation}
The formal expressions for the derivatives in the determinant ${\bf 
D}_{i}$ 
are the same 
as given in Eqs. (B.15) -- (B.17)
of Ref. \cite{Moretto2}. However, the derivatives of the gap 
$\bar{\Delta}$ 
entering in these expressions are more complicate due to Eq. 
(\ref{MBCSgap}).
They are obtained here as
\begin{equation}
\frac{\partial\bar{\Delta}}{\partial{\alpha}}=
\frac{\sum_{j}\Omega_{j}\bigg\{
\frac{\sqrt{2a_{j}}}{\beta}\frac{\bar{\Delta}^{2}}{E_{j}^{2}}-
(\epsilon_{j}-\bar{\mu})[(\epsilon_{j}-\bar{\mu})c_{j}+\bar{\Delta}(a_{j}-b_{j})]\bigg\}}
{\frac{2}{G}-\beta\sum_{j}\Omega_{j}[E_{j}^{2}b_{j}+\bar{\Delta}^{2}(a_{j}-b_{j})]-
\bar{\Delta}\sum_{j}\Omega_{j}(\epsilon_{j}-\bar{\mu})(\beta 
c_{j}+\frac{\sqrt{2a_{j}}}{E_{j}^{2}})}~,
\label{dDda}
\end{equation}
\[
\frac{\partial\bar{\Delta}}{\partial\beta}=
\]
\begin{equation}
\frac{\sum_{j}\Omega_{j}\bigg\{\bar{\Delta}[(\epsilon_{j}-\bar{\mu})(\epsilon_{j}a_{j}-\bar{\mu} 
b_{j})
+\bar{\Delta}^{2}a_{j}]-\frac{\bar{\mu}}{\beta}\bar{\Delta}^{2}
\frac{\sqrt{2a_{j}}}{E_{j}^{2}}+(\epsilon_{j}-\bar{\mu})[(\epsilon_{j}-\bar{\mu})
\epsilon_{j}
+\bar{\Delta}^{2}]c_{j}\bigg\}}
{\frac{2}{G}-\beta\sum_{j}\Omega_{j}[E_{j}^{2}b_{j}+
\bar{\Delta}^{2}(a_{j}-b_{j})]-
\bar{\Delta}\sum_{j}\Omega_{j}(\epsilon_{j}-\bar{\mu})(\beta 
c_{j}+\frac{\sqrt{2a_{j}}}{E_{j}^{2}})}~,
\label{dDdb}
\end{equation}
where
\begin{equation}
a_{j}=\frac{{\rm sech}^{2}(z)}{2E_{j}^{2}}~,\hspace{5mm} 
b_{j}=\frac{{\rm tanh}(z)}{\beta E_{j}^{3}}~,\hspace{5mm} 
c_{j}=\frac{{\rm sech}(z){\rm tanh}(z)}{2E_{j}^{2}}~,\hspace{5mm} 
z=\frac{1}{2}\beta E_{j}~.
\label{abc}
\end{equation}
We've just derived the MHFB theory at finite temperature, which 
includes
the quasiparticle-number fluctuation to 
preserve the unitarity of the modified generalized particle-density 
matrix. We've shown that the limit of this MHFB theory 
reproduces the MBCS equations 
obtained previously in Ref. \cite{MRPA,MBCS}. 
For the sake of completeness we give in the Appendix the proof that, 
by using the secondary Bogoliubov transformation (\ref{w}), the 
modified QRPA indeed conserves the Ikeda sum rule.
\section{Analysis of numerical results}
As an illustration for the modified HFB theory at finite temperature, 
we now discuss in detail the results of numerical calculations within 
its limit, the MBCS theory, of the pairing gap, heat capacity, 
level-density parameter, and level density 
for $^{120}$Sn. The single-particle energies $\epsilon_{j}$
used in the calculations 
are obtained within the Woods-Saxon potential at $T=$0. These 
discrete 
neutron and proton spectra include not only bound but also 
quasi-bound 
levels, which span an energy interval from around 
-40 MeV up to around 17 MeV. They include all the major shells up to 
$N (Z) =$ 126 as well as several levels in the next major shell $N 
(Z)=$ 126 -- 184 up to $1k_{17/2}$ orbital. They are 
assumed here to be independent of $T$. 
This assumption is supported by the results of 
temperature-dependent HF calculations, 
which show that for $T\alt$ 5 MeV the variation of the 
single-particle energies 
with 
$T$ is negligible~\cite{Bonche}. 
The value $G_{\nu}=$ 0.13 MeV is adopted for the neutron 
pairing parameter so that 
the gap $\Delta_{\nu}$
for neutrons is about 1.4 MeV at $T=$ 0.    
\subsection{Temperature dependence of pairing gap}
\subsubsection{Open-shell case: Neutron pairing in 
$^{120}$Sn}                                                   
Since the modified gap $\bar{\Delta}$ is a function of $T$, the last 
term $\delta\Delta$ (\ref{dD}) at the r.h.s of Eq. 
(\ref{MBCSgap})                                              
raises a question about the validity of MBCS equation at high 
temperature.
In fact, at first glance, it seems that, 
if the single-particle spectrum is such that $\delta\Delta$ 
is negative and its absolute value is greater than that of the first 
term at the 
r.h.s of Eq. (\ref{MBCSgap}) at a 
certain value of $T$, the gap $\bar{\Delta}$ 
will turn negative and the MBCS approximation breaks down.
In this section we will show that this does not happen in numerical 
calculations using the entire single-particle energy spectrum.
                                                           
Shown in Fig. 1 are the Bogoliubov coefficients 
$u_{j}$, 
$v_{j}$, quasiparticle occupation number $n_{j}$, together with the
combinations $u_{j}v_{j}$, $u_{j}^{2}-v_{j}^{2}$, $1-2n_{j}$, and 
$\delta{\cal N}_{j}$ as functions of single-particle energy 
$\epsilon_{j}$ for neutrons at several temperatures. These
quantities determine the  
behavior of the gap (\ref{MBCSgap}) as a function of 
$T$.  They are rather symmetric functions from both sides 
of the chemical potential $\bar{\mu}$. The latter varies 
weakly around -6 MeV as $T$ increases. 
The product $u_{j}v_{j}$ decreases quickly 
with increasing $T$. At $T=$ 5 MeV, it remains effective only 
in the region of $\pm$5 MeV around $\bar{\mu}$. 
The difference $u_{j}^{2}-v_{j}^{2}$, which 
enters in the thermal part $\delta\Delta$, remains rather 
insensitive to the variation of $T$. In general, the effect of 
pairing on
the Bogoliubov coefficients $u_{j}$ and $v_{j}$ and their 
combinations,
$u_{j}v_{j}$, and $u^{2}_{j}-v^{2}_{j}$, is significant only in the 
region of at most 
$\pm 10$ MeV around $\bar{\mu}$. This situation 
is rather similar to that obtained within 
the BCS theory. However, for the quasiparticle occupation number 
$n_{j}$ and its 
combinations, $1-2n_{j}$, and $\delta{\cal N}_{j}$, the situation is 
different. Here, with increasing $T$, these quantities, although 
having a peak 
near $\bar{\mu}$, spread over the 
whole single-particle spectrum as shown in Fig. 1 (d) 
- (f). For the quantal component $(\Delta_{\rm Q})_{j}$ the 
maximum of $u_{j}v_{j}$ comes always with the minimum of $(1-2n_{j})$ 
near $\bar{\mu}$. Beyond this region the product 
$u_{j}v_{j}(1-2n_{j})$ is small. However, for the thermal-fluctuation 
part of 
the gap, both regions far above and below $\bar{\mu}$ 
are important. This means that, in difference with 
the BCS theory, where one can restrict the calculations with 
valence nucleons on some closed-shell core by renormalizing
the pairing parameter $G_{\nu}$, the calculations for open-shell 
nuclei within the MBCS theory
are necessary to be carried out using the entire single-particle 
spectrum.
                                                                                                                      
This observation is demonstrated on Fig. 2, where 
the partial quantal $(\Delta_{\rm Q})_{j}$ and thermal 
$\delta\Delta_{j}$ gaps are shown as functions of single-particle
energy $\epsilon_{j}$ at several temperatures. 
The quantal part $(\Delta_{\rm Q})_{j}$ is always larger around 
$\bar{\mu}$, but its magnitude quickly decreases as $T$ increases.
On the contrary, the thermal part $\delta\Delta_{j}$
is positive at $\epsilon_{j}<\bar{\mu}$, and negative
at $\epsilon_{j}>\bar{\mu}$. Its absolute value
sharply increases with increasing $T$. In a realistic spectrum  
the number of single-particle levels below $\bar{\mu}$ 
is usually larger than that of those above it. 
In the present example of $^{120}$Sn, within the same energy 
interval of $\sim$ 20 MeV from 
$\bar{\mu}$, the one below $\bar{\mu}$ has 
12, while the one above $\bar{\mu}$ has only 8 single-particle 
levels.  
Therefore, the sum of partial thermal gaps $\delta\Delta_{j}$ has 
more components
in the region below $\bar{\mu}$, where the difference
$v_{j}^{2}-u_{j}^{2}$ is positive. As the result, by summing 
over all single-particle levels weighted over the 
shell degeneracy $\Omega_{j}$, the ensuing thermal 
gap $\delta\Delta$ (\ref{dD}) is always positive.
                                                                
Shown in Fig. 3 are 
the quantal $\Delta_{\rm Q}$ (dashed line) 
and thermal $\delta\Delta$ (dash-dotted line) gaps 
together with the total MBCS gap $\bar{\Delta}$ (thick solid line) 
as functions of $T$. 
The BCS gap is also shown as the dotted line for comparison. It 
collapses at a critical temperature 
$T_{\rm c}\approx$ 0.79 MeV. This value almost coincides 
with the temperature of superfluid - normal phase transition 
estimated for infinite systems, which is about 
0.567$\Delta(T=0)$~\cite{Landau}. On the contrary, within the MBCS 
theory, 
the quantal gap $\Delta_{\rm Q}$  
never collapses, but decreases monotonously with increasing $T$.
The thermal component $\delta\Delta$ increases first with $T$ at 
$T\alt$ 1 MeV, then starts to decrease with increasing $T$ further, 
but still does not vanish even at  $T\sim$ 5 -- 6 MeV. As the result, 
the total 
MBCS gap $\bar{\Delta}$ has a temperature-dependence similar to
that of the quantal gap $\Delta_{\rm Q}$, except for a 
low-temperature region 0.5 MeV $\alt T\alt T_{\rm c}$, where it increases slightly with $T$ because
of the thermal gap $\delta\Delta$. As high $T$, the 
total gap $\bar\Delta$ 
decreases monotonously with increasing $T$. This yields 
a long tail extending up to $T\sim$ 5 - 6 
MeV.                                                                

In order to see how the change of configuration space
affects the calculation of the MBCS gap, we also carried out several 
tests using cut-off spectra.
Examples are shown in Fig. 4. 
The dashed line is the neutron gap obtained in the MBCS calculation  
after removing 
three lowest major shells (up to $N=28$) from the single-particle 
energy spectrum. The calculations are then carried out by putting 
$N=$ 42 particles on the $N=$28 core. 
The balance in the sum over the single-particle levels
is lost with less levels below $\bar{\mu}$ 
participating in the summation. The symmetry of the spectrum 
with respect to $\bar{\mu}$ 
is destroyed. The gap collapses again, but at a much higher 
temperature $T\approx$ 4 MeV, although up to $T\simeq$ 2.5 MeV its 
temperature dependence is almost the same as that obtained
using the entire spectrum.
Removing from the other side of $\bar{\mu}$
two highest levels $1k_{17/2}$ and $1i_{11/2}$ makes 
the reduced spectrum rather 
symmetric again with respect to $\bar{\mu}$.
The balance in summation of $\delta\Delta_{j}$ is restored. 
As the result, the temperature dependence of the gap  
is recovered as shown by the thin solid line. 
However, if one removes further one 
more level, namely the $1j_{15/2}$ one, i.e. the 
reduced space  
consists of only three major shells, 28 -- 50, 50 -- 82, and 82 -- 
126, 
the balance is destroyed 
again with more weight toward the positive values of 
$\delta\Delta_{j}$.
The reduced spectrum now spreads from around -17 MeV up to around 1.6 
MeV, which is strongly asymmetric with respect to $\bar{\mu}$.
In consequence, the high-temperature tail of the gap 
becomes much more enhanced as shown by the dash-dotted line in Fig. 4.
Other tests using a 8-neutron, 20-neutron, and 50-neutron 
cores also show a similar 
feature. In these tests the parameter $G_{\nu}$ is renormalized so as 
to
obtain the same value for $\Delta_{\nu}(T=0)$. 
With such renormalization of $G_{\nu}$ the 
BCS gap always remains the same.
 
These results show the difference in practical calculations within 
the BCS and MBCS 
theories. In the BCS case, the calculation of the gap 
using a closed-shell core with a simple
renormalizition of the pairing parameter $G_{\nu}$ yields the same result 
as that obtained using the entire single-particle energy 
spectrum.
In the MBCS case, the most reliable way is to use the entire 
or as larger as possible single-particle spectrum. 
If using a limited spectrum is unavoidable, 
care should be taken to maintain the
balance in the summation of partial thermal gap $\delta\Delta_{j}$. 
Otherwise, a resulting collapse or an enhanced tail of the 
gap in the high-$T$ region 
would be simply an artifact caused by a limited space.
As a matter of fact, a 
criterion for a good reduction is that the cut-off spectrum should be 
rather 
symmetric with respect to the region where the quantal pairing 
correlations are 
strongest, namely from both sides of the chemical potential, so that 
the 
effect of quasiparticle-number fluctuation is properly taken 
into account (See Figs. 1 (d) - (f) and 2 
(b)). 
It is worth noticing that the limitation of the configuration 
space also yields a wrong behavior 
of the specific heat. This effect is known as the Schottky anomaly, 
according to which the 
specific heat reaches a maximum at a certain temperature and 
decreases 
as temperature increases further~\cite{Schottky}.
\subsubsection{Closed-shell case: Thermally induced pairing 
correlations for protons in $^{120}$Sn}

The MBCS gap equation (\ref{MBCSgap}) also implies that, in principle, 
thermal fluctuations 
can induce pairing correlations even for closed-shell (CS) nuclei. 
However, the situation here is different from that of the open-shell 
nuclei because of a large shell gap between the 
highest occupied (hole) orbital and the lowest empty (particle) 
one, which is about 6 MeV for protons in $^{120}$Sn.
At $T=$ 0 all the orbitals below $\bar{\mu}$ 
are fully occupied ($v_{j_{h}}$=1, $u_{j_{h}}=$ 0, 
$\epsilon_{j_{h}}-\mu<$ 0), while 
those above $\bar{\mu}$ are empty ($v_{j_{h}}$=0, $u_{j_{h}}=$ 1, 
$\epsilon_{j_{p}}-\mu>$ 0). Therefore the 
quantal gap $\Delta_{\rm Q}$ (\ref{DQ}) is always zero.
Pairing 
is so weak that no scattering into the next major shell (particle 
orbitals) is possible. In such situation 
the approximation of the same pairing matrix elements  
may not be extended across a too large shell gap separating hole and 
particle orbitals, especially when $f_{j_{h}}\equiv 1-n_{j_{h}} \gg 
n_{j_{p}}\equiv f_{j_{p}}$, where $f_{j}$ is the single-particle 
occupation number. This restricts
the summation at the r.h.s of Eq. (\ref{MBCSgap}) to be carried 
out at most over only the hole states. 
The MBCS gap $\bar{\Delta}$  
in this case is solely determined by 
the thermal gap $\delta\Delta$ (\ref{dD}) due to 
the quasiparticle-number fluctuation,
namely
\begin{equation}
\bar{\Delta}_{\rm CS}=\delta\Delta_{\rm CS}
\approx G_{\pi}\sum_{j_{h}}\Omega_{j_{h}}
\sqrt{n_{j_{h}}(1-n_{j_{h}})}.
 \label{gapZ}
 \end{equation}
The thermally induced 
gap $\bar{\Delta}_{\pi}$  
for closed-shell 
proton system ($Z=$ 50) in $^{120}$Sn, 
obtained using Eq. (\ref{gapZ}) with the same value of pairing
parameter as that for neutrons,
$G_{\pi}=G_{\nu}$, is plotted as a function of 
temperature in Fig. 5. This figure clearly shows that the
pairing gap for a closed-shell system is different from zero at 
$T\neq$ 0 and increases as $T$ increases. However,
its magnitude, which reaches 
a value of only around 2.6$\times$10$^{-5}$ MeV at $T=$ 5 MeV, 
is practically negligible as compared to $\bar{\Delta}_{\nu}$.
Therefore we will put $\bar{\Delta}_{\pi}$ equal to zero 
in further discussions.
\subsubsection{Comparison between microscopic and macroscopic 
descriptions of thermal fluctuation}
The effect of thermal fluctuations on the pairing gap was first
studied using the Landau macroscopic theory of phase 
transition~\cite{Landau} by Moretto in Ref. \cite{Moretto}.
Within the Landau theory, $\Phi$ (\ref{grandpotMBCS}) is treated as 
a 
function of the independent parameter $\Delta$. 
The probability that the nucleus has any given value of $\Delta$ for 
the pairing gap is 
determined by the isothermal distribution
\begin{equation}
    P(\Delta)\propto{\rm e}^{\Phi(\Delta)}~.
    \label{P}
    \end{equation}
 The averaged gap $\langle\Delta\rangle$ is calculated 
as~\cite{Moretto}
 \begin{equation}
     \langle\Delta\rangle=\frac{\int_{0}^{\infty}\Delta 
     P(\Delta)d\Delta}{\int_{0}^{\infty}P(\Delta)d\Delta}~.
     \label{avegap}
     \end{equation}
This approach does not 
include 
 quantal fluctuations. Therefore, as has been pointed out in Refs. 
 \cite{Goodman2,Landau,Moretto}, at very low temperature or if 
 nonequilibrium states vary too rapidly with time, quantum 
 fluctuations dominate and Eq. (\ref{avegap}) is no longer 
meaningful. 
                    
The probability distribution 
 $P(\Delta)$ (\ref{P}), calculated using the same neutron 
single-particle 
 spectra for $^{120}$Sn and the same pairing parameter $G_{\nu}$, is 
plotted 
 as a function of $\Delta$ at low and high temperatures in 
 Fig. 6 (a) and (b), respectively. At very low temperature, the 
 most probable value, which is the BCS gap, 
 coincides with the averaged one, resulting in a 
 Gaussian-like shape with a peak at the BCS value of 
$\Delta(T=0)\simeq$ 
 1.4 MeV. 
 As $T$ increases, the distribution becomes skewed toward 
 the lower values of $\Delta$. Its maximum, which still corresponds 
to the 
 solution of the BCS equation, moves to lower $\Delta$ 
 and reaches $\Delta=$ 0 at $T=T_{\rm c}$.  
 This is shown in Fig. 6 (a), which is very 
 similar to what obtained before in Fig. 1 of Ref. \cite{Moretto} for 
 a uniform spectrum. 
 As $T$ increases further, the maximum of the 
 distribution still remains 
 at $\Delta=$ 0, while its width  
 continues to increase, 
 showing the increase of thermal fluctuations. At hypothetically high 
 temperatures (Fig. 6 (b))
 the distribution approaches a Gaussian one in the following form
\begin{equation}
      P(\Delta)|_{T\rightarrow\infty}\sim\frac{1}{\sqrt{2\pi}\sigma}
     {\rm 
exp}\bigg(-\frac{\Delta^{2}}{2\sigma^{2}}\bigg)~\hspace{5mm} 
     {\rm with}\hspace{5mm} 
     \sigma=\sqrt{\frac{GT}{2}}~.
     \label{Gaussian}
     \end{equation}
 This behavior means that Eq. (\ref{P}) 
 assumes the effect of thermal fluctuations on the 
 pairing gap to be chaotic at both low and high $T$. By substituting 
Eq. 
 (\ref{Gaussian}) into the r.h.s of Eq. (\ref{avegap}), 
 the integrals can be carried out analytically. The result is
 \begin{equation}
 \langle\Delta\rangle|_{T\rightarrow\infty}=
 \sqrt{\frac{GT}{\pi}}.
 \label{largeT}
 \end{equation}
 This result reveals the increase of the averaged gap 
 $\langle\Delta\rangle$ with increasing $T$ at very high $T$ within
 the Landau theory using the 
 probability function (\ref{P}).
                                                      
The temperature dependence of $\langle\Delta\rangle$ 
for neutrons in $^{120}$Sn is displayed
in Fig. 7 in comparison with the MBCS gap 
$\bar{\Delta}$. 
It is seen from this figure that the agreement between 
the microscopic treatment of thermal fluctuations in the pairing gap
within the MBCS theory and the macroscopic one can be called at best 
qualitative.
The gap does not collapse in both treatments, but while the 
tail of MBCS gap $\bar{\Delta}$ clearly decreases at high $T$ 
with increasing $T$, the
temperature dependence of the 
averaged gap $\langle\Delta\rangle$ remains rather flat, and 
even starts to 
increase slightly with $T$ already at $T>$ 1 MeV 
because of Eq. (\ref{largeT}). This yields a 
$\langle\Delta\rangle$ of about 0.6 MeV even at $T\sim$ 4 -- 5 MeV.
In the low temperature region, 
$\langle\Delta\rangle$ drops at a lower $T\sim$ 0.5 MeV 
as compared to the MBCS gap.
The reason is that the MBCS gap incorporates the microscopic 
interplay 
between the quantal gap $\Delta_{\rm Q}$ and thermal one, 
$\delta\Delta$. At low $T$ the former dominates. The 
gap $\langle\Delta\rangle$ takes into account only thermal 
fluctuations around the most probable value following the distribution
(\ref{P}). The latter assumes equally strong coupling between 
$\Delta$ and all 
the intrinsic degrees of freedom, disregarding quantal effects.
\subsection{Temperature dependence of heat capacity, lelvel-density 
parameter, and level density}
The heat capacity $C$ and inverse level-density parameter $K=A/a$ 
obtained within
the BCS and MBCS theories are shown in Fig. 8 as functions 
of $T$. The heat capacity usually serves as an indicator
for phase transitions. Within the BCS theory, a sharp discontinuity
in $C$ is seen at $T=T_{\rm c}$, where the gap collapses. Together 
with 
the collapse at $T_{\rm c}$ of the BCS gap as an order parameter, 
this behavior of the heat capacity is a 
clear signature of the second-order phase transition~\cite{Landau}.
However, within the MBCS theory, this phase transition is washed out
so that the temperature dependence of the heat capacity is a  
smooth curve with only a slight effect of the bending of the pairing 
gap
in the region 0.5 $\alt T\alt$ 2 MeV. A similar feature of the heat 
capacity has been recently reported for iron isotopes within the
shell-model Monte-Carlo approach~\cite{Liu}. Since both the order 
parameter $\bar{\Delta}$ and the heat capacity are now continuous
functions we can say that no phase transition actually occurs. At 
high $T$ both the MBCS and BCS results approach each 
other.                             
                                                          
The inverse level-density parameter $K=A/a$ obtained 
within the 
MBCS theory is larger than that obtained within the BCS theory at 
$T\alt $ 3 MeV. At higher temperatures both theories predict almost 
the same $K$. Except for the low-temperature region 
(below $T_{\rm c}$), where the Fermi-gas formula (\ref{a-level}) 
is not valid, $K$ increases with increasing $T$ at $T\agt$ 1 MeV, and 
enters the region of the experimentally extracted 
values between 8 $\sim$ 12 MeV at $T\agt$ 2.5 MeV~\cite{aex}.     
At $T\approx$ 1 MeV, the value of $K$ predicted by the MBCS theory is 
around 6 MeV, which is about twice larger than that given by the BCS 
theory.

Shown in Fig. 9 is the logarithm of level density 
$\rho(N, Z)$ (\ref{leveldensity}) as a function of $T$. 
The BCS result shows 
a kink at $T=T_{\rm c}$, while the MBCS result is a smooth curve, 
which increases monotonously as $T$ increases, exposing no signal of 
phase transition. At $T\agt$ 2 MeV both the BCS and MBCS results 
practically coincide.
\section{Conlusions}
This work has derived the modified HFB (MHFB) theory at finite 
temperature, which  
conserves the unitarity relation of the generalized
particle-density matrix. This has been done by including the thermal 
fluctuation of quasiparticle number microscopically in the 
quasiparticle-density matrix. It has been shown that 
the latter can also be obtained by applying 
the secondary Bogoliubov 
transformation discussed in Refs. \cite{MRPA,MBCS}.
The MHFB equations at finite 
temperature have been then derived following the standard variational 
procedure used in Ref. \cite{Goodman1}. 
Its BCS limit yields the modified BCS (MBCS) equations, which 
have been derived previously in Refs. \cite{MRPA,MBCS} using the 
above-mentioned secondary Bololiubov transformation.
Apart from being able to restore the unitarity transformation, 
this secondary transformation helps the
modified QRPA to completely 
restore the Ikeda sum rule for Fermi and Gamow-Teller transitions, 
which has been violated within the renormalized QRPA.

The illustration of the MHFB theory has been presented within the 
MBCS theory by calculating the neutron pairing gap and thermodynamic 
quantities for $^{120}$Sn. 
Detailed analyses of the results obtained show that the calculations
for open-shell nuclei
within the MBCS theory need to be carried out using the entire 
single-particle spectrum, which includes both bound and quasibound 
levels in a large configuration space of about 7 major shells up to 
126 -- 
184 one. When the use of a reduced spectrum is unavoidable, 
the reduction should be 
done symmetrically from both side of the chemical potential 
$\bar{\mu}$ so that the distribution of quasiparticle-occupation 
number can be properly taken into account as it is symmetric 
with respect to $\bar{\mu}$. The MBCS gap decreases monotonously with 
increasing $T$ and does not vanish even at $T\sim$ 5 MeV. The
discontinuity in the BCS heat capacity at 
the critical temperature $T_{\rm c}$ is also competely washed out, 
showing no signature of superfluid-normal 
phase transition. The temperature dependences of 
level density and level-density parameter 
are also smooth. 

The behavior of the MBCS gap as a function of $T$ is found in 
qualitative 
agreement with that given by the macroscopic treatment using 
the Landau theory of phase transitions in the sense that both gaps do 
not collapse at the critical temperature of the BCS superfluid-normal 
phase transition. However, 
quantitative discrepancies between microscopic and 
mascroscopic approaches are evident. In the low-temperature region, 
due to the 
microscopic interplay between quantal and thermal components, the 
MBCS gap starts to decrease at a higher $T$ with increasing $T$ as 
compared to the macroscopically averaged gap $\langle\Delta\rangle$.
At high temperatures $T>$ 2 MeV, the MBCS gap continues to decrease, 
while
$\langle\Delta\rangle$ remains nearly constant and even start to  
increases with increasing $T$.

The MBCS equations also 
shows that thermal fluctuations can induce
a pairing gap even for closed-shell nuclei.  
Results obtained 
using a single-particle space restricted to hole orbitals have shown 
that such a thermally induced 
gap increases with increasing temperature.
However, its magnitude is negligible compared with the gap
in open-shell nuclei. Therefore, it can be safely put to be equal to 
zero at $T\alt$ 5 -- 6 MeV. 

\begin{acknowledgments}
Numerical calculations were performed using an Alpha AXP 
work-station and the Visual-Numeric IMSL libraries 
at the RIKEN Computer Science Laboratory.
\end{acknowledgments}
\appendix
\section{Restoration of Ikeda sum rule within the modified QRPA}
At $T=$ 0, in general, if the quasiparticle correlations are 
significant so  
that the correlated ground state $|\widetilde{0}\rangle$ deviates 
appreciably from the quasiparticle vacuum (\ref{qpvacuum}) or QRPA 
vacuum, the secondary Bogoliubov transformation (\ref{Bogo2}) can be
used to derive a symmetry-conserving theory, which treats the 
ground-state 
correlations within a microscopic and self-consistent framework. In 
this 
case the quasiparticle occupation number $n_{j}$, which characterizes 
the
magnitude of the ground-state correlations, can be evaluated from
the renormalized 
QRPA backward-going amplitudes ${\cal Y}_{jj'}^{(\lambda i)}$ as has 
been
discussed thoroughly in Refs. \cite{MRPA,MBCS,RRPA}.    
In this section we will prove that the modified QRPA theory, 
which has been derived in Ref.~\cite{MRPA} using the secondary 
Bogoliubov 
transformation in the form of Eq. (\ref{w}), indeed conserves the 
Ikeda sum rule.
\vspace{-7mm} 

\subsection{Ikeda sum rule}
The Ikeda sum rule for Fermi ($J=$ 0) and Gamow-Teller ($J=$ 1) 
transitions is defined with 
respect to the ground state $|g.s.\rangle$ of the final nucleus 
$(N,Z)$ as
\begin{equation}
{\cal S}^{-}-{\cal S}^{+}=\sum_{i}|\langle Ji|\beta^{-}|g.s.\rangle|^{2}-
\sum_{i}|\langle Ji|\beta^{+}|g.s.\rangle|^{2}=(2J+1)(N-Z)~,
\label{SmSp}
\end{equation}
where the squared $\beta^{-}$-transition matrix element $|\langle 
Ji|\beta^{-}|g.s.\rangle|^{2}$ 
is calculated as
\begin{equation}
|\langle Ji|\beta^{-}|g.s.\rangle|^{2}=
\bigg|\langle 
Ji|\sum_{j_{\pi}j_{\nu}}q_{j_{\pi}j_{\nu}}[u_{j_{\pi}}v_{j_{\nu}}
A_{j_{\pi}j_{\nu}}^{\dagger}(JM)+u_{j_{\nu}}v_{j_{\pi}}
A_{j_{\pi}j_{\nu}}(J\widetilde{M})]|g.s.\rangle\bigg|^{2}~,
\label{beta-}
\end{equation}
and $\beta^{+}=(\beta^{-})^{\dagger}$. The 
notation ${\cal O}_{J\widetilde{M}}=(-)^{J-M}{\cal O}_{J-M}$ is used 
hereafter, and $q_{j_{\pi}j_{\nu}}$ denotes single-particle matrix 
elements corresponding to the Fermi or Gamow-Teller transition.
The subscripts $\pi$ and $\nu$ denote proton and neutron, 
respectively.
The quasiparticle-pair 
operators $A_{j_{\pi}j_{\nu}}^{\dagger}(JM)$ and 
$A_{j_{\pi}j_{\nu}}(JM)$ are
\begin{equation}
A_{j_{\pi}j_{\nu}}^{\dagger}(JM)=\sum_{m_{\pi}m_{\nu}}\langle 
j_{\pi}m_{\pi}j_{\nu}m_{\nu}|JM\rangle
\alpha_{j_{\pi}m_{\pi}}^{\dagger}
\alpha_{j_{\nu}m_{\nu}}^{\dagger}~,\hspace{5mm} 
A_{jj'}(JM)=[A_{jj'}(JM)^{\dagger}]^{\dagger}~.
\label{A+A}
\end{equation}
Their exact commutation relation is
\[
[A_{j_{\pi}'j_{\nu}'}(J'M'),A_{j_{\pi}j_{\nu}}^{\dagger}(JM)]=
\delta_{JJ'}\delta_{MM'}\delta_{j_{\pi}j_{\pi}'}\delta_{j_{\nu}j_{\nu}'}
\]
\begin{equation}
-\delta_{j_{\pi}j_{\pi}'}\sum_{m_{\pi}m_{\nu}m_{\nu}'}
\langle j_{\pi}m_{\pi}j_{\nu}'m_{\nu}'|J'M'\rangle
\langle j_{\pi}m_{\pi}j_{\nu}m_{\nu}|JM\rangle
\alpha_{j_{\nu}m_{\nu}}^{\dagger}\alpha_{j_{\nu}'m_{\nu}'}
\label{[AA]}
\end{equation}
\[
-\delta_{j_{\nu}j_{\nu}'}\sum_{m_{\pi}m_{\nu}m_{\pi}'}
\langle j_{\pi}'m_{\pi}'j_{\nu}m_{\nu}|J'M'\rangle
\langle j_{\pi}m_{\pi}j_{\nu}m_{\nu}|JM\rangle
\alpha_{j_{\pi}m_{\pi}}^{\dagger}\alpha_{j_{\pi}'m_{\pi}'}~.
\]
\subsection{Fulfillment of Ikeda sum rule within QRPA}
The QRPA treats the excited state $|Ji\rangle$ as a one-phonon state
\begin{equation}
    |Ji\rangle=Q_{JMi}^{\dagger}|{\rm RPA}\rangle~,
\label{JiRPA}
\end{equation}
while the ground state $|{\rm RPA}\rangle$ 
of an even-even nucleus is treated as the
phonon vacuum 
\begin{equation}
    Q_{JMi}|{\rm RPA}\rangle=0~.
    \label{RPAgs}
    \end{equation}
 The $\pi\nu$-phonon operator $Q_{JMi}^{\dagger}$ is defined as
 \begin{equation}
Q_{JMi}^{\dagger}=\sum_{j_{\pi}j_{\nu}}[X_{j_{\pi}j_{\nu}}^{(Ji)}
A_{j_{\pi}j_{\nu}}^{\dagger}(JM)-
Y_{j_{\pi}j_{\nu}}^{(Ji)}
A_{j_{\pi}j_{\nu}}(J\widetilde{M})]~,\hspace{5mm} 
Q_{JMi}=[Q_{JMi}^{\dagger}]^{\dagger}~.
\label{Qpn}
\end{equation}
In order to obtain a set of linear equations with respect to
the amplitudes $X_{j_{\pi}j_{\nu}}^{(Ji)}$ and 
$Y_{j_{\pi}j_{\nu}}^{(Ji)}$, the QRPA assumes the quasiboson 
approximation, which neglects the contribution of the last two terms 
$\sim\alpha^{\dagger}\alpha$ at 
the r.h.s of Eq. (\ref{[AA]}) in the average over the
ground state $|{\rm RPA}\rangle$, i.e.
\begin{equation}
\langle{\rm RPA}|
[A_{j_{\pi}'j_{\nu}'}(J'M'),A_{j_{\pi}j_{\nu}}^{\dagger}(JM)]|{\rm 
RPA}\rangle\approx
\delta_{JJ'}\delta_{MM'}\delta_{j_{\pi}j_{\pi}'}\delta_{j_{\nu}j_{\nu}'}~.
\label{QBA}
\end{equation}
Within this approximation (\ref{QBA}), the condition for the 
$\pi\nu$-phonon operators (\ref{Qpn}) to be bosons, i.e. satisfy the 
commutation relation
\begin{equation}
    \langle{\rm RPA}|[Q_{JMi},Q_{J'M'i'}^{\dagger}]|{\rm 
    RPA}\rangle=\delta_{JJ'}\delta_{MM'}\delta_{ii'}~,
    \label{[QQ]}
    \end{equation}
leads to the following normalization relation for the 
amplitudes  $X_{j_{\pi}j_{\nu}}^{(Ji)}$ and 
$Y_{j_{\pi}j_{\nu}}^{(Ji)}$ 
\begin{equation}
    \sum_{j_{\pi}j_{\nu}}[X_{j_{\pi}j_{\nu}}^{(Ji)}
    X_{j_{\pi}j_{\nu}}^{(J'i')}-Y_{j_{\pi}j_{\nu}}^{(Ji)}
    Y_{j_{\pi}j_{\nu}}^{(J'i')}]=\delta_{JJ'}\delta_{ii'}~,
    \label{normal}
    \end{equation}
Using the inverse transformation of (\ref{Qpn}), one expresses 
$A_{j_{\pi}j_{\nu}}^{\dagger}(JM)$ and $A_{j_{\pi}j_{\nu}}(JM)$ in 
terms of the $\pi\nu$-phonon operators $Q^{\dagger}_{JMi}$ and 
$Q_{JMi}$. Substituting the result into Eq. (\ref{beta-}) and using it
to evaluate the left-hand side (l.h.s.) of Eq. (\ref{SmSp}) we obtain
\[
({\cal S}^{-}-{\cal S}^{+})_{\rm RPA}=
\sum_{i}|\langle{\rm RPA}|Q_{JMi}\beta^{-}|{\rm RPA}\rangle|^{2}-
\sum_{i}|\langle{\rm RPA}|Q_{JMi}\beta^{+}|{\rm RPA}\rangle|^{2}
\]
\begin{equation}
=|\sum_{j_{\pi}j_{\nu}}q_{j_{\pi}j_{\nu}}(u_{j_{\pi}}v_{j_{\nu}}X_{j_{\pi}j_{\nu}}+
v_{j_{\pi}}u_{j_{\nu}}Y_{j_{\pi}j_{\nu}})|^{2}-
|\sum_{j_{\pi}j_{\nu}}q_{j_{\nu}j_{\pi}}(v_{j_{\pi}}u_{j_{\nu}}X_{j_{\pi}j_{\nu}}+
u_{j_{\pi}}v_{j_{\nu}}Y_{j_{\pi}j_{\nu}})|^{2}~
\label{SRPA}
\end{equation}
\[
=\sum_{j_{\pi}j_{\nu}}|q_{j_{\pi}j_{\nu}}|^{2}
(v_{j_{\nu}}^{2}-v_{j_{\pi}}^{2})=
2(2J+1)[\sum_{j_{\nu}}\Omega_{j_{\nu}}v_{j_{\nu}}^{2}-\sum_{j_{\pi}}\Omega_{j_{\pi}}v_{j_{\pi}}^{2}]
=(2J+1)(N-Z)~.
\]
In the above derivation the normalization condition (\ref{normal}) 
and the usual BCS equation for the particle number are used together 
with the property 
$\sum_{j_{i}}|q_{j_{i}j_{k}}|^{2}=2(2J+1)\Omega_{j_{k}}$ 
($i=(\pi,\nu)$, 
$k=(\nu,\pi)$) for the single-particle matrix elements of Fermi and 
Gamow-Teller 
transitions. This derivation shows that the QRPA fulfills the Ikeda 
sum 
rule.
\subsection{Violation of Ikeda sum rule within the renormalized QRPA}
By neglecting the contribution of the two last terms at the r.h.s of 
Eq. (\ref{[AA]}) in the ground state (\ref{RPAgs}), the quasibson 
approximation
(\ref{QBA}) ignores the Pauli principle between the 
quasiparticle-pair 
operators (\ref{A+A}). This causes the collapse of the QRPA at a 
certain critical value of the interaction parameter, where the 
solution of the QRPA equations becomes imaginary. The renormalized
QRPA has been proposed as a method to cure this 
inconsistency~\cite{Hara,RRPA,Krmpotic}.

This approach assumes that, instead of the quasiboson approximation (\ref{QBA}), 
the following commutation relation 
holds in the average over the correlated ground state 
$|\widetilde{\rm RPA}\rangle$
\begin{equation}
\langle\widetilde{\rm RPA}|
[A_{j_{\pi}'j_{\nu}'}(J'M'),A_{j_{\pi}j_{\nu}}^{\dagger}(JM)]|
\widetilde{\rm RPA}\rangle=
\delta_{JJ'}\delta_{MM'}\delta_{j_{\pi}j_{\pi}'}\delta_{j_{\nu}j_{\nu}'}D_{j_{\pi}j_{\nu}}~,
\label{RRPA}
\end{equation}
where
\begin{equation}
    D_{j_{\pi}j_{\nu}}=1-n_{j_{\pi}}-n_{j_{\nu}}~,\hspace{5mm} 
    n_{j}=\frac{1}{2\Omega_{j}}\langle\widetilde{\rm 
    RPA}|\alpha_{jm}^{\dagger}\alpha_{jm}|\widetilde{\rm 
    RPA}\rangle\neq 0~.
    \label{Dpn}
    \end{equation}
This means that the renormalized QRPA takes into account the 
contribution of the diagonal elements of the last two terms at the 
r.h.s of Eq. (\ref{[AA]}) in the correlated ground state 
$|\widetilde{\rm RPA}\rangle$.

The renormalized $\pi\nu$-phonon operators ${\cal Q}_{JMi}^{\dagger}$ 
and ${\cal Q}_{JMi}$ are introduced as
\begin{equation}
{\cal Q}_{JMi}^{\dagger}=
\sum_{j_{\pi}j_{\nu}}\frac{1}{\sqrt{D_{j_{\pi}j_{\nu}}}}[{\cal 
X}_{j_{\pi}j_{\nu}}^{(Ji)}
A_{j_{\pi}j_{\nu}}^{\dagger}(JM)-
{\cal Y}_{j_{\pi}j_{\nu}}^{(Ji)}
A_{j_{\pi}j_{\nu}}(J\widetilde{M})]~,\hspace{5mm} 
{\cal Q}_{JMi}=[{\cal Q}_{JMi}^{\dagger}]^{\dagger}~.
\label{Qcalpn}
\end{equation}
The correlated ground state $|\widetilde{\rm RPA}\rangle$ is defined 
as the vacuum with respect to the renormalized
phonon operators, i.e.
\begin{equation}
    {\cal Q}_{JMi}|\widetilde{\rm RPA}\rangle=0~.
    \label{RRPAgs}
    \end{equation}
Because of Eq. (\ref{RRPA}), these renormalized $\pi\nu$ phonon 
operators
satisfy the boson commutation relation 
\begin{equation}
    \langle\widetilde{\rm RPA}|[{\cal Q}_{JMi},{\cal 
    Q}_{J'M'i'}^{\dagger}]|\widetilde{\rm 
    RPA}\rangle=\delta_{JJ'}\delta_{MM'}\delta_{ii'}~
    \label{[QcalQcal]}
    \end{equation}
in the correlated ground state $|\widetilde{\rm RPA}\rangle$ 
provided their ${\cal X}$ and ${\cal Y}$ amplitudes satisfy the same 
normalization condition as in the QRPA, i.e.
\begin{equation}
    \sum_{j_{\pi}j_{\nu}}[{\cal X}_{j_{\pi}j_{\nu}}^{(Ji)}
    {\cal X}_{j_{\pi}j_{\nu}}^{(J'i')}-{\cal 
Y}_{j_{\pi}j_{\nu}}^{(Ji)}
    {\cal Y}_{j_{\pi}j_{\nu}}^{(J'i')}]=\delta_{JJ'}\delta_{ii'}~,
    \label{Rnormal}
    \end{equation}
The phonon energies, ${\cal X}_{j_{\pi}j_{\nu}}^{(Ji)}$ and
${\cal Y}_{j_{\pi}j_{\nu}}^{(Ji)}$ amplitudes
are found by solving the nonlinear QRPA-like equations, whose 
submatrices contains the factor $D_{j_{\pi}j_{\nu}}$.
The latter is found from the equation~\cite{MRPA}
\begin{equation}
{D}_{jj'}=1
-\sum_{Ji}(J+1/2)\sum_{j''}\{{D}_{jj''}\frac{[{\cal Y}_{jj''}^{(J 
i)}]^{2}}{\Omega_{j}}
+
{D}_{j''j'}\frac{[{\cal Y}_{j''j'}^{(J i)}]^{2}}{\Omega_{j'}}
\}~,
\label{Djj}
\end{equation}
The presence of the factor $D_{j_{\pi}j_{\nu}}$
makes the solution of the renormalized QRPA always real as
the interaction strength is reduced by this factor so that the 
collapse
is avoided.     
However, the inverse transformation of (\ref{Qcalpn}) now becomes
 \begin{equation}
 A_{j_{\pi}j_{\nu}}^{\dagger}(JM)=\sqrt{D_{j_{\pi}j_{\nu}}}\sum_{JMi}
 [{\cal X}_{j_{\pi}j_{\nu}}^{(Ji)}
{\cal Q}_{JMi}^{\dagger}+
{\cal Y}_{j_{\pi}j_{\nu}}^{(Ji)}
{\cal Q}_{J\widetilde{M}i}]~,\hspace{5mm} A_{j_{\pi}j_{\nu}}(JM)=
[A_{j_{\pi}j_{\nu}}^{\dagger}(JM)]^{\dagger}~.
\label{inverseQcal}
\end{equation}
Using Eq. (\ref{inverseQcal}) to evaluate the quantity $S^{-}-S^{+}$ 
in the same way as in the derivation (\ref{SRPA}), one finds
\[
({\cal S}^{-}-{\cal S}^{+})_{\rm RRPA}=
\sum_{i}|\langle\widetilde{\rm RPA}|{\cal 
Q}_{JMi}\beta^{-}|\widetilde{\rm RPA}\rangle|^{2}-
\sum_{i}|\langle\widetilde{\rm RPA}|{\cal 
Q}_{JMi}\beta^{+}|\widetilde{\rm RPA}\rangle|^{2}
\]
\begin{equation}
=\sum_{j_{\nu}j_{\pi}}D_{j_{\pi}j_{\nu}}
|q_{j_{\pi}j_{\nu}}|^{2}(v_{j_{\nu}}^{2}-v_{j_{\pi}}^{2})~.
\label{SRRPA}
\end{equation}
This quantity is smaller than $(2J+1)(N-Z)$ since 
$D_{j_{\pi}j_{\nu}}<$ 1. Hence, Eq. (\ref{SRRPA}) shows that the 
renormalized QRPA violates the Ikeda sum rule.

We notice that, although the renormalized QRPA takes into account 
Eq. (\ref{RRPA}), it neglects the following commutation relation 
between the scattering-quasiparticle pairs
\begin{equation}
\langle\widetilde{\rm RPA}|
[B_{j_{\pi}'j_{\nu}'}^{\dagger}(J'M'),B_{j_{\pi}j_{\nu}}(JM)]|
\widetilde{\rm RPA}\rangle=
\delta_{JJ'}\delta_{MM'}\delta_{j_{\pi}j_{\pi}'}\delta_{j_{\nu}j_{\nu}'}(n_{j_{\nu}}-n_{j_{\pi}})~,
\label{[BB]}
\end{equation} 
where 
\begin{equation}
B_{j_{\pi}j_{\nu}}(JM)=-\sum_{m_{\pi}m_{\nu}}\langle 
j_{\pi}m_{\pi}j_{\nu}m_{\nu}|JM\rangle\alpha_{j_{\pi}m_{\pi}}^{\dagger}
\alpha_{j_{\nu}\widetilde{m_{\nu}}}~,
\label{B}
\end{equation}
The omission of the contribution of scattering-quasiparticle 
operators $B_{j_{\pi}'j_{\nu}'}^{\dagger}(J'M')$ and 
$B_{j_{\pi}j_{\nu}}(JM)$ is the source that leads to 
the underestimation of the quantity ${\cal S}^{-}-{\cal S}^{+}$
within the renormalized QRPA.
\subsection{Restoration of Ikeda sum rule within the modified QRPA}
The modified QRPA makes a further step by taking into account the 
effects 
of ground-state correlations on the quasiparticle and collective 
excitations. This has been realized in Ref. \cite{MRPA} using 
the secondary Bogoliubov transformation (\ref{w}), where $n_{j}$ is 
the quasiparticle-occupation number in the new correlated ground state
$|\overline{\rm RPA}\rangle$:
\begin{equation}
n_{j}=\langle\overline{\rm 
RPA}|\alpha_{jm}^{\dagger}\alpha_{jm}|\overline{\rm RPA}\rangle\neq 
0~.
\label{nj}
\end{equation}
The modified QRPA phonon operators are introduced as
\begin{equation}
\bar{Q}_{JMi}^{\dagger}=\sum_{j_{\pi}j_{\nu}}[\bar{X}_{j_{\pi}j_{\nu}}^{(Ji)}
\bar{A}_{j_{\pi}j_{\nu}}^{\dagger}(JM)-
\bar{Y}_{j_{\pi}j_{\nu}}^{(Ji)}
\bar{A}_{j_{\pi}j_{\nu}}(J\widetilde{M})]~,\hspace{5mm} 
\bar{Q}_{JMi}=[\bar{Q}_{JMi}^{\dagger}]^{\dagger}~,
\label{QMRPA}
\end{equation} 
where $\bar{A}_{j_{\pi}j_{\nu}}^{\dagger}(JM)$ and 
$\bar{A}_{j_{\pi}j_{\nu}}(JM)$
are the creation and destruction operators of a 
modified-quasiparticle 
pair
\begin{equation}
\bar{A}_{j_{\pi}j_{\nu}}^{\dagger}(JM)=\sum_{m_{\pi}m_{\nu}}\langle 
j_{\pi}m_{\pi}j_{\nu}m_{\nu}|JM\rangle\bar{\alpha}_{j_{\pi}m_{\pi}}^{\dagger}
\bar{\alpha}_{j_{\nu}m_{\nu}}^{\dagger}~,\hspace{5mm} 
\bar{A}_{jj'}(JM)=[\bar{A}_{jj'}(JM)^{\dagger}]^{\dagger}~.
\label{AAbar}
\end{equation}
The new ground state $|\overline{\rm RPA}\rangle$ is defined as the 
vacuum 
for the modified phonon operator, i.e. 
\begin{equation}
\bar{Q}|\overline{\rm RPA}\rangle= 0~.
\label{MQRPAgs}
\end{equation}
The transformation from the single-particle operators 
$a_{jm}^{\dagger}$ and $a_{jm}$ to the modified quasiparticle opertors
$\bar{\alpha}_{jm}^{\dagger}$ and $\bar{\alpha}_{jm}$ is obtained 
after 
successively applying the usual (\ref{uv}) and secondary (\ref{w}) 
Bogoliubov transformations, and has the form similar to the inverse 
transformation of (\ref{uv})
\begin{equation}
a_{jm}^{\dagger}=\bar{u}_{j}\bar{\alpha}_{jm}^{\dagger}+(-)^{j-m}\bar{v}_{j}
\bar{\alpha}_{j-m}~,\hspace{5mm} 
(-)^{j-m}a_{j-m}=(-)^{j-m}\bar{u}_{j}\bar{\alpha}_{j-m}
-\bar{v}_{j}\bar{\alpha}_{jm}^{\dagger}~,
\label{ajm}
\end{equation}
where
\begin{equation}
\bar{u}_{j}=u_{j}\sqrt{1-n_{j}}+v_{j}\sqrt{n_{j}}~,
\hspace{5mm} 
\bar{v}_{j}=v_{j}\sqrt{1-n_{j}}-u_{j}\sqrt{n_{j}}~.
\label{ubarvbar}
\end{equation}
Using the secondary Bogoliubov transformation (\ref{w}) to express
${A}_{j_{\pi}j_{\nu}}^{\dagger}(JM)$ and 
${A}_{j_{\pi}j_{\nu}}(JM)$ in terms of 
$\bar{A}_{j_{\pi}j_{\nu}}^{\dagger}(JM)$, 
$\bar{A}_{j_{\pi}j_{\nu}}(JM)$, 
$\bar{B}_{j_{\pi}j_{\nu}}^{\dagger}(JM)$, 
and $\bar{B}_{j_{\pi}j_{\nu}}(JM)$ (See Eqs. (9) and (10) of Ref. 
\cite{MRPA}), we find that Eqs. (\ref{RRPA}) and (\ref{[BB]}) hold in 
the correlated ground state (\ref{MQRPAgs}) if 
$\bar{B}_{j_{\pi}j_{\nu}}^{\dagger}(JM)$, 
and $\bar{B}_{j_{\pi}j_{\nu}}(JM)$ commute, while
$\bar{A}_{j_{\pi}j_{\nu}}^{\dagger}(JM)$, 
$\bar{A}_{j_{\pi}j_{\nu}}(JM)$ 
obey the same commutation relation (\ref{QBA}) with respect to the 
correlated 
ground state $|\overline{\rm RPA}\rangle$. Therefore the set of 
equations
to define the energy and amplitudes $\bar{X}_{j_{\pi}j_{\nu}}^{(Ji)}$ 
and $\bar{Y}_{j_{\pi}j_{\nu}}^{(Ji)}$ 
of the modified $\pi\nu$-phonon excitation have the same form as 
of the usual QRPA ones. This set of equations is called 
the modified QRPA equations. The amplitudes 
$\bar{X}_{j_{\pi}j_{\nu}}^{(Ji)}$ 
and $\bar{Y}_{j_{\pi}j_{\nu}}^{(Ji)}$ obey the same normalization 
condition as in Eq. (\ref{normal}), namely
\begin{equation}
    \sum_{j_{\pi}j_{\nu}}[\bar{X}_{j_{\pi}j_{\nu}}^{(Ji)}
    \bar{X}_{j_{\pi}j_{\nu}}^{(J'i')}-\bar{Y}_{j_{\pi}j_{\nu}}^{(Ji)}
    \bar{Y}_{j_{\pi}j_{\nu}}^{(J'i')}]=\delta_{JJ'}\delta_{ii'}~,
    \label{normalM}
    \end{equation}
The set of modified QRPA equations 
should be solve simultaneously with the normalization condition 
(\ref{normalM}) and  
the equation for $n_{j}$. The latter is evaluated from Eq. 
(\ref{Djj}) as 
\begin{equation}
    n_{j}=\sum_{Ji}\frac{2J+1}{2j+1}\sum_{j'}{D}_{jj'}[{\cal 
Y}_{jj'}^{(J 
i)}]^{2}~,
\label{njMRPA}
\end{equation}
where ${\cal Y}_{jj'}^{(J i)}$ is expressed in terms of 
${\bar{X}}_{jj'}^{(J i)}$ and ${\bar{Y}}_{jj'}^{(J i)}$ as [See Eq. 
(16) of Ref. \cite{MRPA})]
\begin{equation}
    {\cal Y}_{jj'}^{(J i)}={\bar{X}}_{jj'}^{(J i)}
    \sqrt{(1-n_{j})(1-n_{j'})}+\bar{Y}_{jj'}^{(J i)}
    \sqrt{n_{j}n_{j'}}~.
    \label{YcalXbarYbar}
    \end{equation}
Because of the renormalization factors $\sqrt{(1-n_{j})(1-n_{j'})}$
and $\sqrt{n_{j}n_{j'}}$ the modified QRPA can avoid the collapse
in a way similar to that of the renormalized QRPA.
Using the same secondary Bogoliubov transformation (\ref{w}), it is 
easy to see that the modified QRPA phonon operators contain 
all the operators ${A}_{j_{\pi}j_{\nu}}^{\dagger}(JM)$ and 
${A}_{j_{\pi}j_{\nu}}(JM)$, ${B}_{j_{\pi}j_{\nu}}^{\dagger}(JM)$, 
and ${B}_{j_{\pi}j_{\nu}}(JM)$ in its definition [See Eq. (13) of 
Ref. \cite{MRPA}]. This feature allows the modified QRPA to restore 
the Ikeda sum rule as shown below.

The quantity $S^{-}-S^{+}$ is calculated within the modified QRPA as
\[
({\cal S}^{-}-{\cal S}^{+})_{\rm MRPA}=
\sum_{i}|\langle\overline{\rm RPA}|\bar{Q}_{JMi}\beta^{-}|\overline
{\rm RPA}\rangle|^{2}-
\sum_{i}|\langle\overline{\rm RPA}|\bar{Q}_{JMi}\beta^{+}|\overline
{\rm RPA}\rangle|^{2}
\]
\begin{equation}
=|\sum_{j_{\pi}j_{\nu}}q_{j_{\pi}j_{\nu}}(\bar{u}_{j_{\pi}}\bar{v}_{j_{\nu}}\bar{X}_{j_{\pi}j_{\nu}}+
\bar{v}_{j_{\pi}}\bar{u}_{j_{\nu}}\bar{Y}_{j_{\pi}j_{\nu}})|^{2}-
|\sum_{j_{\pi}j_{\nu}}q_{j_{\nu}j_{\pi}}(\bar{v}_{j_{\pi}}\bar{u}_{j_{\nu}}\bar{X}_{j_{\pi}j_{\nu}}+
\bar{u}_{j_{\pi}}\bar{v}_{j_{\nu}}\bar{Y}_{j_{\pi}j_{\nu}})|^{2}~
\label{SMPA}
\end{equation}
\[
=\sum_{j_{\pi}j_{\nu}}|q_{j_{\pi}j_{\nu}}|^{2}
(\bar{v}_{j_{\nu}}^{2}-\bar{v}_{j_{\pi}}^{2})=
2(2J+1)[\sum_{j_{\nu}}\Omega_{j_{\nu}}\bar{v}_{j_{\nu}}^{2}-\sum_{j_{\pi}}\Omega_{j_{\pi}}
\bar{v}_{j_{\pi}}^{2}]=(2J+1)(N-Z)~,
\]
making use of the normalization condition (\ref{normalM}), 
Eq. (\ref{ubarvbar}) for $\bar{v}_{j}$ in terms of $u_{j}$ 
and $v_{j}$, and MBCS Eq. (\ref{MBCSN}) for particle number. We've 
just proved 
that the modified QRPA restores the Ikeda sum rule. 


\newpage
{\bf FIGURE CAPTIONS}
\vspace{5mm} 

{\bf Fig. 1} Occupation probabilities within MBCS theory 
as functions of single-particle 
energies $\epsilon_{j}$ for neutrons at $T=$ 0.4, 1, 3, and 5 MeV.
(A thicker line corresponds to a higher $T$ as indicated in panels (a) 
and (d)). Panel (a) shows the Bogoliubov coefficients
$u_{j}$ (dotted lines) and $v_{j}$ (solid lines). 
Pannels (b) and (c) show the  product $u_{j}v_{j}$ and
the difference $u^{2}_{j}-v^{2}_{j}$, respectively.
Panel (d) shows the quasiparticle occupation number $n_{j}$.
Panels (e) and (f) show the factor $(1-2n_{j})$ and 
$\delta{\cal N}_{j}\equiv\sqrt{n_{j}(1-n_{j})}$, respectively.
The open circles marked on the lines at $T=$ 5 MeV in (a) and (d) 
correspond to the positions of single-particle levels.
\vspace{5mm}

{\bf Fig. 2} Partial quantal $(\Delta_{\rm Q})_{j}$ and thermal 
$\delta\Delta_{j}$ gaps as functions of 
single-particle 
energies $\epsilon_{j}$ for neutrons at $T=$ 0.4, 1, 3, and 5 MeV.
A thicker line corresponds to a higher temperature as indicated in 
panel (a).
\vspace{5mm}

{\bf Fig. 3}
Neutron pairing gap as a function of temperature $T$.
The thick solid line represents the MBCS gap $\bar{\Delta}$. Its two 
components, the quantal gap $\Delta_{\rm Q}$ and thermal gap 
$\delta\Delta$, are shown by dashed and dash-dotted lines, 
respectively.
The BCS gap is shown by the dotted line.
\vspace{5mm}

{\bf Fig. 4} Neutron pairing gap as a function of temperature $T$.
The thick solid line represents the MBCS gap $\bar{\Delta}$ obtained
using the entire single-particle spectrum,
as in Fig. 3. 
The dashed line shows the result obtained using an $N=28$ core. 
The thin solid line represents the result 
obtained using the same core and removing 
two highest levels $1k_{17/2}$ and $1j_{11/2}$. 
The dash-dotted line is the result obtained using
the same core and removing 
three highest levels, $1k_{17/2}$, $1j_{11/2}$, 
and $1j_{15/2}$.
The dotted line shows the BCS gap. 
\vspace{5mm}

{\bf Fig. 5} Thermally induced pairing gap $\bar{\Delta}_{\pi}$ 
for protons in $^{120}$Sn as a function of $T$. 
\vspace{5mm}

{\bf Fig. 6} Probability distribution $P(\Delta)$ for $^{120}$Sn 
as a function of gap parameter $\Delta$ 
at different temperatures shown by the numbers
(in MeV) next to the thick solid curves.
The dashed curves in (b) represent the Gaussian distribution 
(\ref{Gaussian}) at 
various temperatures shown by the italic numbers (in MeV) 
next to these curves.
\vspace{5mm}

{\bf Fig. 7} Neutron pairing gap for $^{120}$Sn 
as a function of $T$ within 
the MBCS theory and the macroscopic treatment following the Landau 
theory of phase transitions. The thick and thin 
lines show the gap $\bar\Delta$ and the averaged gap 
$\langle\Delta\rangle$, respectively.
\vspace{5mm}

{\bf Fig. 8} Heat capacity (a) and the inverse level-density parameter 
$K=A/a$ (b) as 
functions of temperature for $^{120}$Sn. The thick solid lines
denote results obtained within the MBCS theory. The dotted lines
show the BCS results. 
\vspace{5mm}

{\bf Fig. 9} Logarithm of level density as a function of $T$
for $^{120}$Sn. The notation is as in Fig. 8.


\begin{thebibliography}{99}
\bibitem[1]{Goodman1}A. L. Goodman, Nucl. Phys. A {\bf 352}, 30 (1981).
\bibitem[2]{Tanabe}K. Tanabe, K. Sugawara-Tanabe, and H.J. Mang, Nucl. 
Phys. A {\bf 357}, 20 (1981).
\bibitem[3]{LN}H.J. Lipkin, Ann. of. Phys. {\bf 31}, 525 (1960); Y. 
Nogami 
and I.J. Zucker, Nucl. Phys. {\bf 60}, 203 (1964); Y. Nogami, Phys. 
Lett. {\bf 15}, 335 (1965); J.F. Goodfellow and Y. Nogami, 
Can. J. Phys. {\bf 44}, 1321 (1966); 
H.C. Pradhan, Y. Nogami, and J. Law, Nucl. Phys. A {\bf 201}.
357 (1973).
\bibitem[4]{Sheikh}J.A. Sheikh, P. Ring, E. Lopes, and R. Rossignoli,
Phys. Rev. C {\bf 66}, 044318 (2002).
\bibitem[5]{RingSchuck}P. Ring and P. Schuck, The Nuclear Many Body 
Problems (Springer, Berlin, 2000).
\bibitem[6]{Quentin}N. Pillet, P. Quentin, J. Libert, Nucl. Phys. A {\bf 
697}, 141 (2002).
\bibitem[7]{Egido}J.L. Egido, Phys. Rev. Lett. {\bf 61}, 767 (1988).
\bibitem[8]{DangZ}N. Dinh Dang, Z. Phys. A {\bf 335}, 253 (1990).
\bibitem[9]{Goodman2}A.L. Goodman, Phys. Rev. C {\bf 29}, 1887 (1984).
\bibitem[10]{Sandu}O. Civitarese, G.G. Dussel, and R.P.J. Prerazzo, Nucl. 
Phys. A {\bf 404}, 15 (1983); \\
N. Dinh Dang, J. Phys. G {\bf 11}, L125 (1985); 
N. Dinh Dang and N. Zuy Thang, J. Phys. G {\bf 14}, 
1471 (1988); N. Sandulescu, O. Civitarese, and R.J. Liotta, Phys. Rev.
C {\bf 61}, 044317 (2000).
\bibitem[11]{Landau}L.D. Landau and E.M. Lifshitz, Course of Theoretical
Physics, Vol. 5: Statistical Physics (Moscow, Nauka, 1964).
\bibitem[12]{Moretto}L.G. Moretto, Phys. Lett. B {\bf 40}, 1 (1972). 
\bibitem[13]{MRPA}N. Dinh Dang and V. Zelevinsky, Phys. Rev. C {\bf 64}, 
064319 (2001); Ibid. {\bf 65}, 069903(E) (2002).
\bibitem[14]{MBCS}N. Dinh Dang and A. Arima, Phys. Rev. C {\bf 67}, 
014304 (2003). 
\bibitem[15]{DangRing}N.D. Dang, P. Ring, and R. Rossignoli, Phys. Rev. C
{\bf 47}, 606 (1993).
\bibitem[16]{Rossignoli}R. Rossignoli, P. Ring, and N.D. Dang, 
Phys. Lett. B {\bf 297}, 9 (1992).
\bibitem[17]{Liu}S. Liu and Y. Alhassid, Phys. Rev. Lett. {\bf 87}, 
022501 (2001).
\bibitem[18]{Ze} 
V. Zelevinsky, B.A. Brown, N. Frazier, and M. Horoi,
Phys. Rep. {\bf 276}, 85 (1996).
\bibitem[19]{Volya}A. Volya, B.A. Brown, and V. Zelevinsky, Phys. Lett. B 
{\bf 509}, 37 (2001).
\bibitem[20]{Ikeda}K. Ikeda, Prog. Theor. Phys. {\bf 31}, 431 (1964).
\bibitem[21]{Hara}K. Hara, Prog. Theor. Phys. {\bf 32}, 88 (1964); K. 
Ikeda, 
T. Udagawa, and H. Yamamura, {\it ibid.} {\bf 33}, 22 (1965); 
D.J. Rowe, Phys. Rev. {\bf 175}, 1293 (1968); 
Rev. Mod. Phys. {\bf 40}, 153 (1968); Nucl. Phys. {\bf A107}, 99 
(1968); 
A. Klein, R.M. Dreizler, and R.E. Johnson, Phys. Rev. 
{\bf 171}, 1216 (1968); 
P. Schuck and S. Ethofer, Nucl. Phys. {\bf A212}, 269 
(1973); J. Dukelsky and P. Schuck, {\it ibid.} {\bf A512}, 446 (1990).
\bibitem[22]{RRPA}F. Catara, N. Dinh Dang, and M. Sambataro, Nucl. Phys. 
{\bf A 579}, 1 (1994).
\bibitem[23]{Krmpotic}F. Krmpoti\'{c}, E.J.V. de Passos, D.S. Delion, J. 
Dukelsky, and P. Schuck, Nucl. Phys. {\bf A 637}, 295 (1998).
J.G. Hirsch, P.O. Hess, and O. Civitarese, Phys Rev. {\bf C 56}, 199 
(1997).
\bibitem[24]{Civi}J.G. Hirsch, P.O. Hess, and O. Civitarese, Phys. Rev. C
{\bf 54}, 1976 (1996).
\bibitem[25]{ERRPA}N.D. Dang and A. Arima, Phys. Rev. C {\bf 62}, 024303 
(2000).
\bibitem[26]{res}S. Stoica and H.V. Klapdor-Kleingrothaus, Eur. Phys. J. 
A {\bf 9}, 345 (2000); Phys. Rev. C {\bf 63}, 064304 (2001), Nucl. 
Phys. A {\bf 694}, 269 (2001); D.S. Delion, J. Dukelsky, and P. 
Schuck, Phys. Rev. C {\bf 55}, 2340 (1997); V. Rodin and A. Faessler, 
Phys. Rev. C {\bf 66}, 051303 (R) (2002).
\bibitem[27]{Zubarev}D.N. Zubarev, Soviet Physics Uspekhi {\bf 3}, 320 
(1960) [Usp. Fiz. Nauk. {\bf 71}, 71 (1960)].
\bibitem[28]{Tanabe2}K. Tanabe and K. Sugawara-Tanabe, Phys. Lett. B 
{\bf 247}, 202 (1990).
\bibitem[29]{TFD}Y. Takahashi and H. Umezawa, Collective Phenomena, 
{\bf 2}, 55 (1975);
H. Umezawa, H. Matsumoto, and M. Tachiki, {Thermo field
dynamics and condensed states} (North-Holland, Amsterdam, 1982).
\bibitem[30]{Entro}K. Tanabe and N. Dinh Dang, Phys. Rev. C {\bf 62}, 024310 
(2000).
\bibitem[31]{Sano}M. Sano and S. Yamazaki, Prog. Theor. Phys. {\bf 29}, 
397 (1963).
\bibitem[32]{Moretto2}L.G. Moretto, Nucl. Phys. A {\bf 182}, 641 (1972).
\bibitem[33]{Behkami}A.N. Behkami and J.R. Huizenga, Nucl. Phys. A {\bf 
217}, 78 (1973).
\bibitem[34]{Bonche}M. Brack and P. Quentin, Phys. Lett. B {\bf 52}, 159 
(1974); P. Bonche, S. Levit, and D. Vautherin, Nucl. Phys. 
{\bf A 427}, 278 (1984).
\bibitem[35]{Schottky}O. Civitarese, G.G. Dussel, and A.P. Zuker, Phys. 
Rev. 
C {\bf 40}, 2900 (1989); J. Dukelsky, A. Poves, and J. Retamosa, 
Phys. 
Rev. C {\bf 44}, 2872 (1991); O. Civitarese and M. Schvellinger, 
Phys. 
Rev. C {\bf 49}, 1976 (1994).
\bibitem[36]{aex}G. Nebia et al., Phys. Lett. B {\bf 176}, 20 (1986); K. 
Hagel et al., Nucl. Phys. A {\bf 486}, 429 (1988); M. Gonin at el., 
Phys. Lett. B {\bf 217}, 406 (1989); A. Chibi et al., Phys. Rec. C 
{\bf 43}, 666 (1991).
\end{thebibliography}
\end{document}